\documentclass[aps,prd,reprint,longbibliography,nofootinbib,superscriptaddress,floatfix]{revtex4-2}

\usepackage{slashed}
\usepackage{verbatim}
\usepackage[T1]{fontenc}
\usepackage{mathbbol}
\usepackage[normalem]{ulem}
\usepackage[english]{babel}
\usepackage{orcidlink}
\usepackage{physics}
\usepackage{subfig}
\usepackage{dcolumn}
\usepackage{tensor}
\usepackage{comment}
\usepackage{placeins}
\usepackage{graphicx,overpic,mathtools}
\usepackage{amsthm,amsmath,amssymb,mathrsfs}
\usepackage{braket,bm,bbm,setspace}
\usepackage{booktabs}
\usepackage{cancel}
\usepackage{xargs}
\usepackage{ragged2e}
\usepackage[normalem]{ulem}
\usepackage[dvipsnames,table,xcdraw]{xcolor}

\usepackage{hyperref}
\hypersetup{
    colorlinks=true,
    linkcolor=RoyalBlue,
    citecolor=ForestGreen,
    urlcolor=RoyalBlue}

\usepackage[nameinlink,capitalise]{cleveref}

\def\d{\mathrm{d}}

\newcommand{\uib}{%
Departament de F\'isica, Universitat de les Illes Balears,
IAC3~--~IEEC, Crta.\ Valldemossa km~7.5, E-07122 Palma, Spain}
\newcommand{\GSSI}{%
Gran Sasso Science Institute (GSSI), I-67100 L'Aquila, Italy}
\newcommand{\GranSasso}{%
INFN, Laboratori Nazionali del Gran Sasso, I-67100 Assergi, Italy}

\begin{document}

\title{Shaping black hole resonances I:\\ Black hole ringdown as a spectral filtering process}

\author{Alejandro Svyatkovskyy Kholyavka\orcidlink{0009-0004-1688-6018}}
\email{alejandro.svyatkovskyy@uib.cat}
\affiliation{\uib}

\author{Jose Antonio Le\'on Vega\orcidlink{0009-0002-7382-5137}}
\email{jose-antonio.leon1@estudiant.uib.cat}
\affiliation{\uib}

\author{Samuel G\'omez G\'omez}
\email{samuel.gomez@uib.cat}
\affiliation{\uib}

\author{Xisco Jim\'enez Forteza\orcidlink{0000-0002-8158-5009}}
\email{f.jimenez@uib.es}
\affiliation{\uib}

\author{Sayak Datta\orcidlink{0000-0002-4774-0298}}
\email{sayak.datta@gssi.it}
\affiliation{\GSSI}
\affiliation{\GranSasso}

\begin{abstract}
The ringdown of a perturbed black hole (BH) can be described as a superposition of quasinormal modes (QNMs), whose frequencies are determined by the spacetime geometry while their amplitudes depend also on the perturbing source. However, the physical mechanism governing mode excitation remains unclear and is typically treated on a case by case basis.
In this work, we show that QNM excitation is governed by a simple spectral rule: each mode is excited according to the Fourier content of the perturbation evaluated at its characteristic frequency. This result follows from the factorization of the excitation coefficients and establishes a direct, quantitative connection between the spectral properties of the perturbation and the resulting ringdown amplitudes.
To make this mechanism explicit and controllable, we construct localized perturbations with independently tunable spectral bandwidth and carrier frequency. We demonstrate analytically and numerically that BHs act as resonant spectral filters.
We show analytically that the excitation amplitude of each mode equals the weighted spatial Fourier transform of the initial data evaluated at wave number $k\sim\omega_n$ so that the filter selectively excites modes whose frequencies lie within the spectral support of the perturbation while suppressing others. Consequently, the excitation is maximized when the dominant perturbation frequency lies close to the real part of the QNM frequency, and we validate this at the percent level with fits to time-domain numerical evolutions. To robustly perform these fits, we have developed a new fitting algorithm, \texttt{QNMToolkit}, which performs ringdown fits over large ensembles of sliding time-domain windows and quantifies the resulting fitting variance.
\end{abstract}

\preprint{TBD}

\maketitle
\setlength{\parskip}{0.5em}

\section{Introduction}
\label{sec:intro}

The ringdown phase of a perturbed black hole (BH) provides a direct probe of the spacetime geometry in the strong-field regime of general relativity. Following a perturbation, the system undergoes a transient evolution described by a superposition of damped oscillations known as quasinormal modes (QNMs), whose complex frequencies are determined entirely by the underlying spacetime geometry~\cite{vishveshwara:1970zz,1970PhRvD...1.2870V, chandrasekhar,teukolsky1,teukolsky2,press:1971wr}. This makes QNMs a cornerstone of gravitational-wave (GW) astronomy, where they enable precision tests of gravity and BH spectroscopy~\cite{dreyer:2002mx,detweiler:1980gk,berti:2005gp,gossan:2011ha}. Observational tests of the ringdown spectrum of BHs have already been reported by the LIGO--Virgo--KAGRA collaboration~\cite{ligoscientific:2019fpa,LIGOScientific:2025wao,isi2019,Carullo:2019flq,Abbott:2020jks,LIGOScientific:2025rid,LIGOScientific:2025wao} while  third generation observatories will provide a significant improvement \cite{Evans:2021gyd,ET:2025xjr}.

From the perspective of wave propagation, BH perturbations can be understood as a scattering problem in an effective potential~\cite{press:1971wr,teukolsky:1974yv,ferrari:1984zz,leaver:1985ax,Schutz:1986gp}. The QNMs arise as solutions satisfying purely ingoing boundary conditions at the horizon and purely outgoing conditions at infinity, defining a non-Hermitian eigenvalue problem~\cite{leaver:1985ax,chandrasekhar}. They correspond to poles of the Green's Function (GF) in the complex frequency plane, in direct analogy with resonances of open quantum systems \cite{Ching:1998mxl,Gamow:1928zz,Siegert:1939zz,2015RPPh...78k4001R,2011nhqm.book.....M,Kokkotas:1999bd,Konoplya:2011qq}. The observed ringdown signal therefore reflects how the source couples to these poles --- a spectral overlap problem that is the central subject of this work.

While the QNM frequencies are fully determined by the background spacetime, their excitation amplitudes depend on the perturbation and are encoded in quasi-normal mode excitation coefficients (QNECs)~\cite{Kubota:2025hjk,Motohashi:2025,berti:2007fi,Zhang:2013ksa,Oshita:2021iyn,Lo:2025njp,DellaRocca:2025zbe}. Although these coefficients can be computed within the GF formalism, their physical interpretation remains opaque. Most existing studies focus on specific perturbations, which obscures the underlying excitation mechanism and limits predictive power when the source is varied. Moreover, QNM excitations become even less transparent in full numerical relativity (NR) binary simulations, where they must either be inferred through fits to the NR data~\cite{Finch:2021iip,Finch:2022ynt,london:2014cma,Forteza:2022tgq,forteza2020,Forteza:2021wfq,Giesler:2024hcr,Mitman:2025hgy,Cheung:2023vki} or remain implicitly encoded in current waveform models~\cite{khan:2015jqa,cotesta:2018fcv,Estelles:2020twz,Varma:2019csw,Ramos-Buades:2023clm,Garcia-Quiros:2020qpx,Yoo:2023lqk}.
Using linear perturbation theory (LPT), we show in this work that QNM excitation obeys a spectral selection rule: each mode responds only to the Fourier content of the perturbation evaluated at its characteristic frequency. 

We demonstrate that the overlap function $T_n$, governing the QNEC, is a \emph{weighted spatial Fourier transform} of the initial data (ID), evaluated at wave number $k\sim\omega_n$. This exhibits that each mode samples the perturbation spectrum around its own characteristic frequency. The weighting encodes the spacetime response through the QNM wavefunction. Excitation is therefore determined by the spectral content of an effective, spacetime-filtered source. In this picture, the BH acts as a resonant filter, sampling the perturbation spectrum independently at each pole frequency. 
Modes whose frequencies lie outside the spectral support are suppressed, while those aligned with spectral peaks are resonantly enhanced. The width of the perturbation controls the spectral bandwidth, while an oscillatory modulation shifts the spectral power to a chosen frequency, enabling selective excitation of individual modes. Building on the implicit structure of the GF formalism~\cite{leaver:1985ax,Andersson:1995zk,Berti:2006wq} we make this connection explicit, validate it quantitatively, and develop it into a predictive framework for controlling QNM excitation. Our fits to the numerical data are performed using a new numerical algorithm \texttt{QNMToolkit} that constructs a large ensemble of sliding time-domain windows with variable sizes for each waveform. Each window therefore provides an independent estimation of the ringdown amplitudes. The ensemble allows us to agnostically quantify the variance arising from the ambiguity in the choice of the fit starting time, contamination from prompt-response and tail contributions, and power leakage from higher overtones. The algorithm can be found in~\cite{QNMToolkit2026}.

Although derived within the LPT, this framework provides a useful interpretation of ringdown in fully nonlinear scenarios. In binary BH mergers, the effective perturbation is dynamically generated during the coalescence and is not known \emph{a priori}. As a result, the relative excitation of QNMs in NR waveforms cannot be predicted analytically.
From the perspective developed here, the merger can instead be viewed as generating an effective perturbation with a characteristic spectral content. The final BH then filters them according to its QNM spectrum. In this sense, the spectral filtering picture provides a simple and predictive interpretation of mode excitation, capturing the essential features of the effective perturbation generated during the merger
~\cite{pretorius_2005,campanelli:2006fy,baker:2005vv,berti:2007fi}. 

The paper is organized as follows. Section~\ref{sec:formalism} introduces the perturbation formalism and the spectral overlap interpretation of QNM excitation. Section~\ref{sec:id} defines the tunable ID and their spectral properties. Section~\ref{sec:analytical} presents the key analytical result for the excitation amplitude. Section~\ref{sec:numerical} describes the numerical framework. Section~\ref{sec:results} presents the results, including waveform dependence on spectral parameters, excitation coefficients, numerical validation, and source location effects. Section~\ref{sec:multipoles} further analyzes the multipolar structure of the spectral filtering mechanism. Section~\ref{sec:conclusion} concludes. Throughout this work we employ the $\ell=2$ multipole, except in section~\ref{sec:multipoles} and Appendix~\ref{app:multipoles}. We also make use of a geometrized unit system $G=M=c=1$.

\section{Black Hole Response as a Spectral Overlap Problem}
\label{sec:formalism}

Linear gravitational perturbations of a Schwarzschild BH with line element
\begin{equation}
ds^2 = -f(r)\,\d t^2 + f(r)^{-1}\d r^2 + r^2\,\d\Omega^2
\end{equation}
are described by the Regge--Wheeler (RW) master equations~\cite{Regge:1957td}. Owing to the spherical symmetry of the background, perturbations decompose into axial (odd-parity) and polar (even-parity) sectors~\cite{Zerilli:1970se}. Due to the isospectrality of the two sectors~\cite{Chandrasekhar:1984siy}, we focus on the axial sector. However, there exists a transformation relating the axial and polar sectors~\cite{chandrasekhar:1975zza,Zhang:2013ksa}. The perturbation reduces to a single gauge-invariant master variable satisfying
\begin{equation}
\frac{\partial^2\Psi}{\partial r_\star^2} - \frac{\partial^2\Psi}{\partial t^2} - V_\ell(r)\,\Psi = S\big(r_\star, t\big)\,,
\label{eq:wave}
\end{equation}
where $\d r / \d r_\star = f(r) = 1 - 2M/r$ defines the tortoise coordinate $r_\star$. For axial gravitational perturbations, the effective potential is
\begin{equation}
V_{\ell}(r) = f(r)\Bigg(\frac{\ell\,\big(\ell+1\big)}{r^2} - \frac{6M}{r^3}\Bigg)\,,
\label{eq:potential}
\end{equation}
which vanishes at the horizon and at spatial infinity, and peaks near the photon sphere at $r = 3M$.  BH's response to the source $S(r_\star,t)$ can be determined from the GF of the system.
\subsection{Green's function solution and QNM poles}
Following standard practice, we construct the GF from the homogeneous solutions of Eq.~\eqref{eq:wave} subject to appropriate boundary conditions. Since $V_\ell\to 0$ as $r_\ast\to\pm\infty$, the radial equation for the Fourier-transformed variable $\widetilde{\Psi}(r_\ast,\omega)$ admits plane-wave solutions with the asymptotic behavior
\begin{align}
\psi_{\rm in} \big(r_\star,\omega\big) &\sim \begin{cases}
e^{-i\omega r_\star} & r_\star \to -\infty \\[0.5em]
A_{\rm in}e^{-i\omega r_\star} + A_{\rm out}e^{+i\omega r_\star} & r_\star \to +\infty
\end{cases}
\label{eq:psi_in}\\[1em]
\psi_{\rm up}\big(r_\star,\omega\big) &\sim \begin{cases}
B_{\rm in}e^{-i\omega r_\star} + B_{\rm out}e^{+i\omega r_\star} & r_\star \to -\infty \\[0.5em]
e^{+i\omega r_\star} & r_\star \to +\infty
\end{cases}
\label{eq:psi_up}
\end{align}
where $A_{{\rm in/out}}(\omega)$ and $B_{{\rm in/out}}(\omega)$ are the scattering coefficients.
For a nonvanishing source, Eq.~\eqref{eq:wave} is solved in the frequency domain using the GF formalism. The frequency-domain solution reads
\begin{equation}
\hat\Psi\big(r_\star,\omega\big) = \int_{-\infty}^{+\infty} \hat{G}\big(r_\star, r'_\star;\omega\big)\, \mathcal{I}\big(\omega, r'_\star\big)\,\d r'_\star\,,
\label{eq:gf_solution}
\end{equation}
where the GF is
\begin{equation}
G\big(r_\star,r'_\star;\omega\big) = \frac{\psi_{\rm in}\big(r_<,\omega\big)\,\psi_{\rm up}\big(r_>,\omega\big)}{\mathcal{W}(\omega)}\,,
\end{equation}
with $r_< = \min\big(r_\star, r'_\star\big)$ and $r_> = \max\big(r_\star,r'_\star\big)$, and the Wronskian
\begin{equation}
\mathcal{W} = \psi_{\rm in}\,\psi'_{\rm up} - \psi_{\rm up}\,\psi'_{\rm in} = 2i\omega A_{\rm in}(\omega)\,.
\label{eq:wronskian}
\end{equation}
The source term
\begin{equation}
\mathcal{I}\big(r_\star, \omega\big) = \tilde{S}\big(\omega, r_\star\big) - i\omega\,\Psi\big|_{t=0} + \partial_t\Psi\big|_{t=0}\,,
\label{eq:source}
\end{equation}
encodes both the Fourier-transformed external forcing $\widetilde{S}$ and the initial state of the field. The QNMs are defined by imposing $A_{\rm in}(\omega_n) = 0$, which are the poles of the GF. Using the residue theorem, the time-domain solution in the asymptotic limit $r_\star \to \infty$ for a given $\ell$ mode is
\begin{equation}
\Psi\big( r_\star ,t \big) = \sum_n C_n\, e^{-i\omega_n(t - r_\star)}\,.
\label{eq:ringdown}
\end{equation}
Here $\omega_n$ denotes the complex QNM frequencies, $\omega_n = \omega_n^{\rm Re} - i\,\omega_n^{\rm Im}$, with $\omega_n^{\rm Im} > 0$ encoding the damping rate. 

Eq. \eqref{eq:ringdown} is a superposition of outgoing damped oscillations --- the ringdown signal observed at infinity. The time-domain solution is obtained by the inverse Fourier transform 
\begin{equation}
\Psi\big(r_\star,t\big)=\int_{-\infty}^{\infty} \hat\Psi\big(r_\star,\omega\big)\,e^{-i\omega t}\,\d\omega\,.
\end{equation}
Near a pole 
\begin{equation}
A_{\rm in}(\omega) = \cancel{A_{\mathrm{in}}(\omega_{\mathrm{n}})}+ \frac{\d A_{\rm in}}{\d\omega}\Bigg|_{\omega_n} \big(\omega - \omega_n\big) + \cdots\,,
\end{equation}
so that the residue theorem picks out the QNM contributions and the time-domain ringdown is given by Eq.~\eqref{eq:ringdown}. 
\subsection{Excitation as weighted Fourier transform}
From the representation in Eq.~\eqref{eq:ringdown}, $C_n$ admits a natural factorization that separates spacetime properties from source dependence. The excitation amplitude is fully determined by $B_n$~\cite{Andersson:1995zk,Berti:2006wq}, that depends only on the background spacetime
\begin{equation}
B_n = \frac{A_{\rm out}\big(\omega_n\big)}{2\omega_n A'_{\rm in}\big(\omega_n\big)}\,,
\end{equation}
and the quantity
\begin{equation}
T_n = \int_{-\infty}^{\infty} \frac{\psi_{\rm in}\big(r'_\star, \omega_n\big)\, \mathcal{I}\big(r'_\star, \omega_n\big)}{A_{\rm out}\big(\omega_n\big)}\,\d r'_\star\,,
\label{eq:tn}
\end{equation}
for an ID with compact support  entirely located far from the observer~\cite{Berti:2006wq}. $T_n$ is the overlap of the source with the ingoing solution: it quantifies how efficiently the perturbation excites each QNM. 
The excitation amplitude of each mode is therefore
\begin{equation}
C_n = B_n\,T_n\,.
\label{eq:cn}
\end{equation}
We refer to $B_n$ as the quasinormal excitation factor (QNEF), which depends only on the background spacetime, and to $T_n$ as the (spectral) overlap between the perturbation and the $n-$th mode, encoding the source dependence. The full amplitude $C_n$ which determines the contribution of each mode to the ringdown, will be referred to as the QNEC.
Eq.~\eqref{eq:tn} shows that the overlap is a spectral transform of the source weighted by the QNM wavefunction. The kernel $\psi_{\rm in}(r_\star,\omega_n)$ encodes the spatial structure of the mode and the scattering properties of the potential. Factoring out the asymptotic plane wave behavior of the QNM wavefunction
\begin{equation}
\psi_{\rm in}\big(r_\star,\omega_n\big) \equiv A_{\rm out}\big(\omega_n\big)\,e^{i\omega_n r_\star}\, W_n\big(r_\star\big)\,,
\end{equation}
the overlap simplifies as
\begin{equation}
\label{eq:weighted_FT_Tn}
T_n = \int \mathcal{I}\big(r_\star',\omega_n\big)\,e^{i\omega_nr_\star'}\,W_n\big(r_\star'\big)\,dr_\star'\,.
\end{equation}
QNM excitation amplitude, therefore, is a \emph{weighted Fourier transform} of the source evaluated at the QNM frequency. The weighting function $W_n(r_\star)$ encodes the spatial structure of the mode and the scattering properties of the spacetime, thereby defining an effective source
\begin{equation}
\mathcal{I}_{\rm eff}(r_\star)=\mathcal{I}(r_\star)\,W_n(r_\star)\,.
\end{equation}
The overlap is therefore the spectral projection of this effective source. The BH does not sample the perturbation uniformly in space, but filters it according to the QNM wavefunction. At large distance (asymptotic limit) $W_n\to 1$, and overlap reduces to standard Fourier transform evaluated at QNM frequencies. However, for generic perturbations the asymptotic approximation breaks down, and the overlap becomes weighted by the QNM wavefunction.

\begin{figure*}[!ht]
\centering
\includegraphics[width=0.49\linewidth]{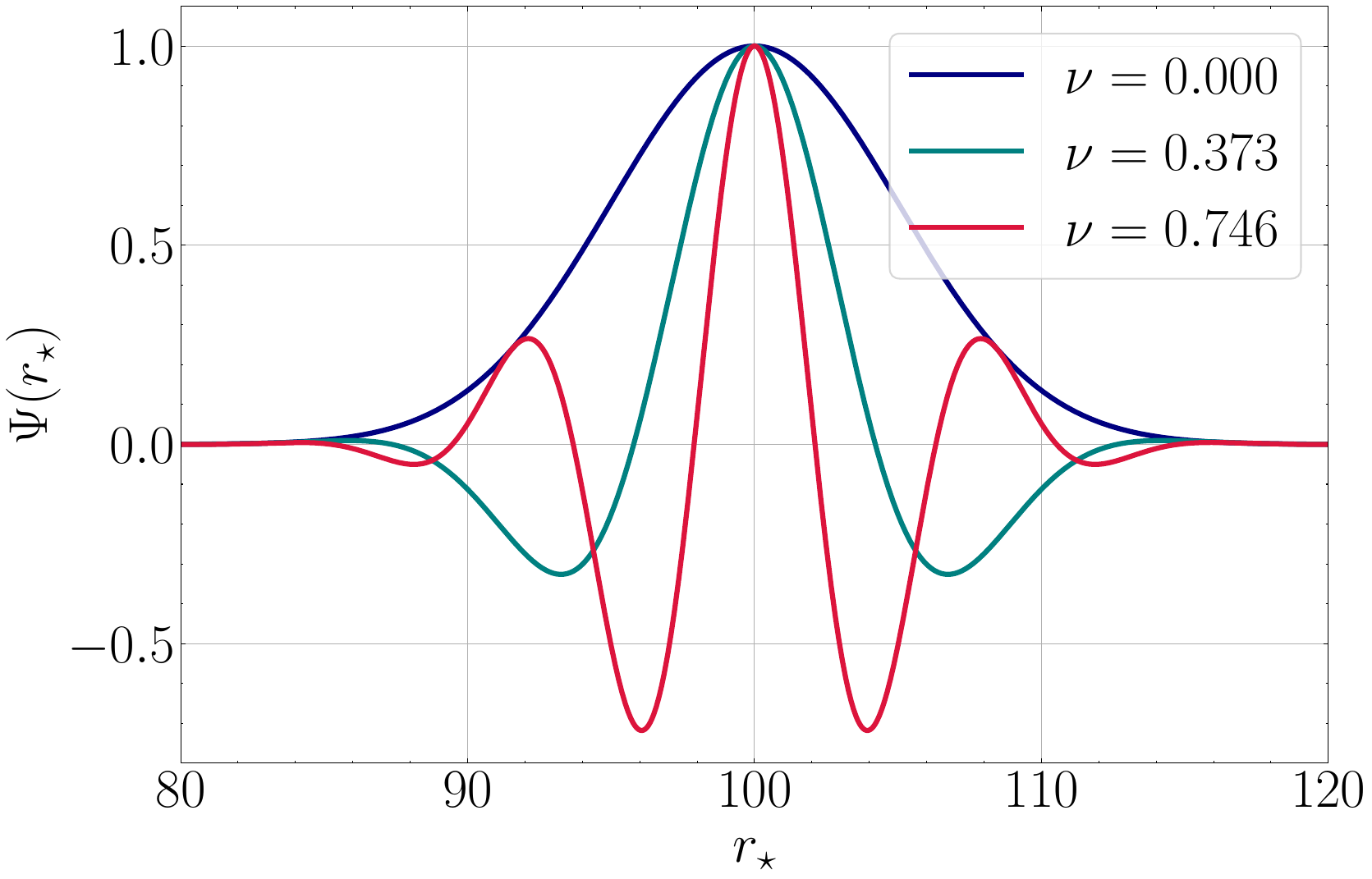}
\includegraphics[width=0.49\linewidth]{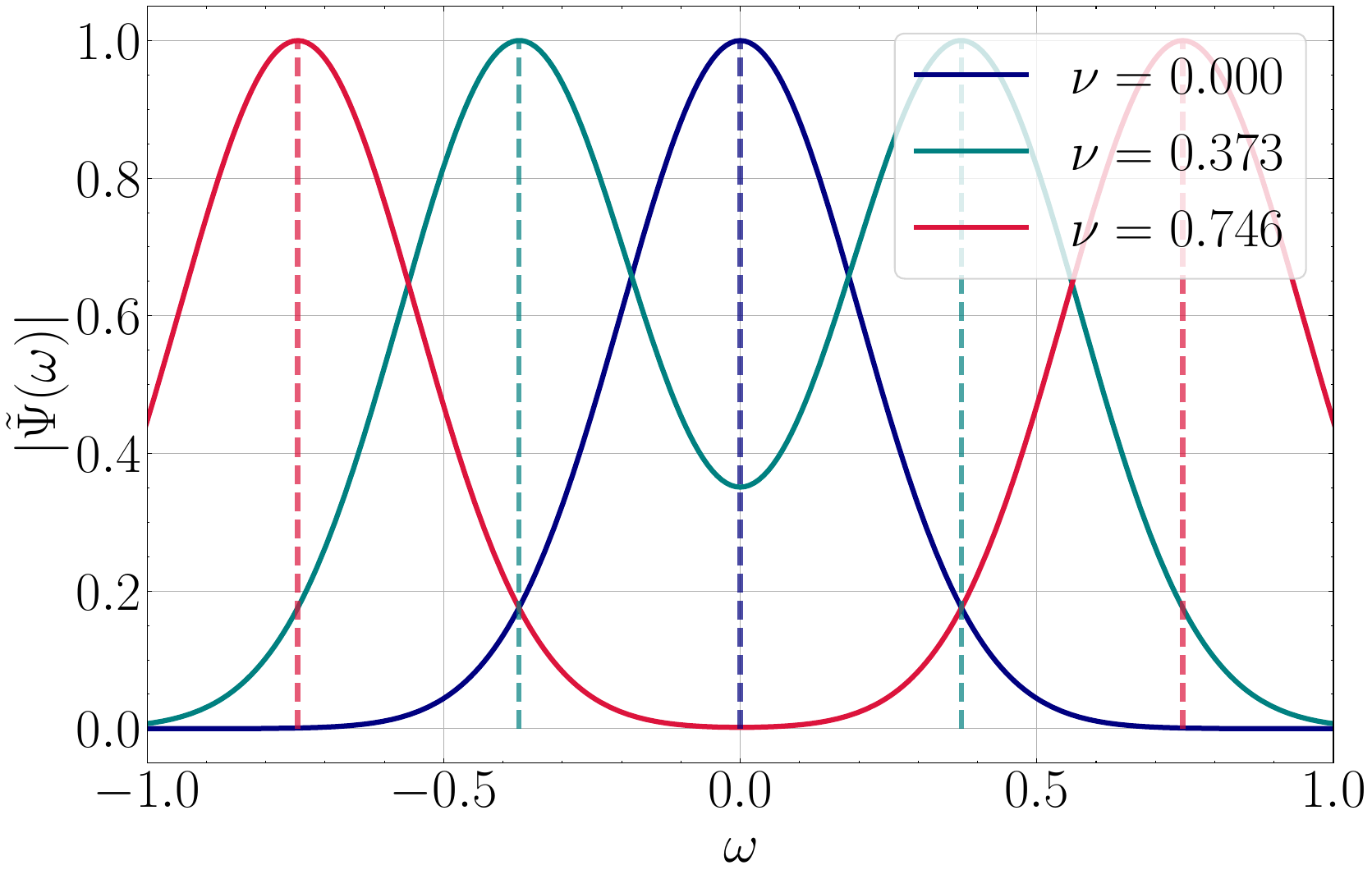}
\includegraphics[width=0.49\linewidth]{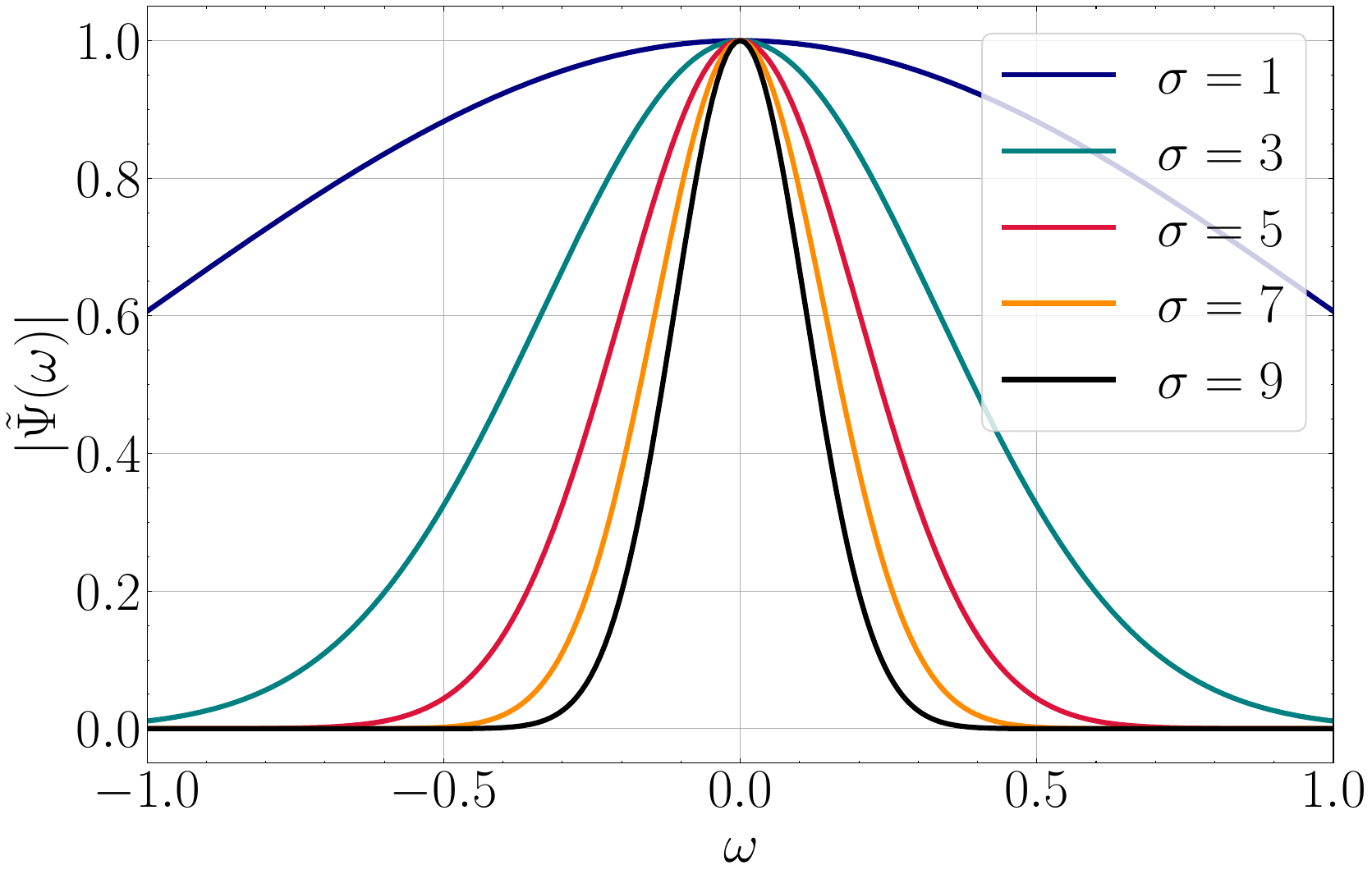}
\includegraphics[width=0.49\linewidth]{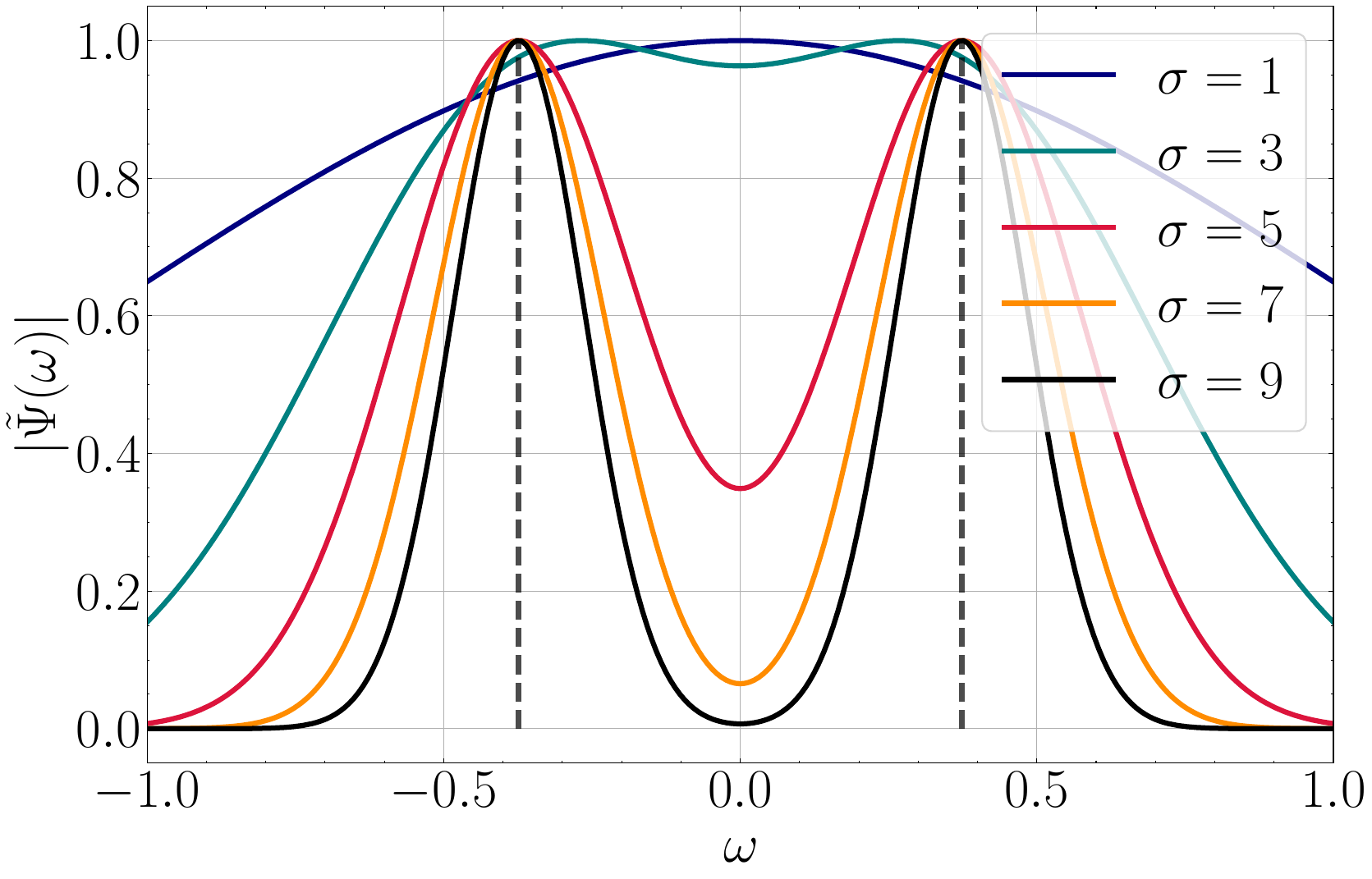}
\caption{\justifying
ID and their spatial Fourier spectra. \textbf{Top left:} Spatial profiles $\Psi(r_\star)$ of the localized perturbations. The Gaussian envelope, characterized by width $\sigma$ and centered at $r_0$, sets the spatial localization, while an oscillatory modulation with frequency $\nu$ introduces a controllable driving frequency. \textbf{Top right:} Corresponding spatial Fourier spectra $|\tilde{\Psi}(\omega)|$. The bandwidth scales as $\Delta\omega \sim 1/\sigma$, while the oscillatory modulation shifts the spectral power toward frequencies $\omega \simeq \pm \nu$. \textbf{Bottom left:} Dependence on $\sigma$ for the pure Gaussian case ($\nu = 0$). Increasing $\sigma$ narrows the spectrum around $\omega = 0$, while the limit $\sigma \to 0$ approaches a flat spectrum. \textbf{Bottom right:} Same as bottom left, but for an oscillatory Gaussian with $\nu = 0.373 \approx\omega_{20}^{\rm Re}$. The spectral structure depends on $\alpha = \sigma \nu$: for $\alpha < 1$ the spectrum is single-peaked near $\omega = 0$, while for $\alpha > 1$ two peaks develop near $\omega = \pm \nu$.}
 \label{fig:combined_ID}
\end{figure*}
%
The amplitude and phase of each mode in the ringdown are therefore independently controllable by tuning the perturbation spectrum. A perturbation whose spectrum has a peak near $\omega_n^{\rm Re}$ will strongly excite mode $n$; one whose spectrum is concentrated far from $\omega_n^{\rm Re}$ will suppress it. In this sense, the BH acts as a bank of resonant filters. Each QNM responds independently to the Fourier content of the source at its own pole frequency through the overlap. All subsequent analysis in this paper reduces to understanding and controlling $T_n$ through the spectral design of the ID. In the following sections we construct families of ID that allow us to control this overlap and thereby selectively excite different modes with the required precision, establishing a direct link between the spectral properties of the perturbation and the excitation of QNMs. 

\section{Spectral Control via Tunable Initial Data}
\label{sec:id}

To probe the spectral filtering mechanism, we construct families of localized ID with independently tunable bandwidth and carrier frequency. We use simple Gaussian IDs as broadband probes, and oscillatory Gaussian pulses for introducing a well-defined driving frequency. This separation of bandwidth and frequency enables selective excitation of individual modes and provides direct, quantitative access to the resonant structure of the spacetime.
We solve the master equations numerically and using GF integrals also find approximate analytical expressions. The analytic approach allows us to quantify BH amplitude resonances without fitting, thereby avoiding the typical issues related to overfitting and numerical resolution errors~\cite{bhagwat:2016ntk,forteza2020,baibhav}.

The structure of the considered localized oscillatory Gaussian pulse is
\begin{equation}
\Psi(r_\star)\Big|_{t=0} = A\,\exp\!\left[-\frac{\big(r_\star - r_0\big)^2}{2\sigma^2}\right]\cos\!\Big[\nu\,\big(r_\star - r_0\big)\Big]\,,
\label{eq:id}
\end{equation}
where $A$ is the amplitude, $r_0$ is the pulse center, $\sigma$ is the characteristic spatial width, and $\nu$ is the carrier frequency. Setting $\nu = 0$ recovers a pure Gaussian. The initial perturbation is placed at $r_0 \gg 3M$, well outside the photon sphere, so that the pulse propagates naturally toward the BH and excites the QNMs rather than forcing an excitation at the potential barrier. 

Throughout this work we fix the amplitude $A=1$, which defines a specific normalization class where the pointwise amplitude of the perturbation is held constant. Since the RW equation is linear, the excitation coefficients scale linearly with $A$, and this choice does not affect relative mode excitation or the structure of the spectral overlap. However, if $A$ is allowed to depend on the parameters $(\sigma,\nu)$, the resulting excitation landscape is modified. We comment on this below.

Upper panel of Fig.~\ref{fig:combined_ID} shows representative spatial profiles (left) and their Fourier spectra (right) for varying $\nu$ at fixed $\sigma = 5$. In the lower panel we show the impact of the spatial width $\sigma$. In the left panel we consider the pure Gaussian, setting $\nu=0$, while in the right panel we show the spectrum for different values of $\sigma$ and $\nu=0.373 \approx\omega^{\rm Re}_{20}$.

The spatial Fourier spectrum of~Eq.~\eqref{eq:id} is $\sqrt{2\pi}A\,\sigma\tilde\Psi(\omega) $ with
\begin{equation}
\tilde\Psi(\omega) = 
\,e^{i\omega r_0}\,e^{-\frac{\sigma^2}{2}\big(\omega^2 + \nu^2\big)}\,\cosh\big(\sigma^2\omega\nu\big)\,.
\label{eq:fourier_id}
\end{equation}

The parameters of this construction admit a direct interpretation in the frequency domain, allowing us to treat them as independent spectral controls. The three parameters $\sigma$, $\nu$, and $r_0$ act as independent spectral controls. Freely propagating waves satisfy $\omega=k$ in the asymptotic region. The spatial modulation frequency $\nu$ therefore directly sets the temporal frequency content delivered to the BH.

The width $\sigma$ determines the spectral bandwidth: the Fourier spectrum has width $\Delta\omega \sim 1/\sigma$, so spatially narrow pulses (small $\sigma$) are spectrally broad while spatially wide pulses are spectrally narrow and concentrated near low frequencies for pure Gaussian, as demonstrated in Fig.~\ref{fig:combined_ID}. In the limit $\sigma \to 0$, the pulse $\tilde{\Psi}(\omega)\rightarrow 1$. In the opposite limit, large $\sigma$ progressively concentrates spectral power near $\omega \sim 0$ for pure Gaussian, eliminating overlap with QNM frequencies and eventually suppressing all ringdown in favor of the late-time tail~\cite{tfm_alex}.

The oscillatory modulation frequency $\nu$ shifts the dominant spectral content toward $\pm\nu$ (for $\sigma\nu \gg 1$), enabling selective frequency targeting independently of $\sigma$. The case $\nu = 0$ recovers the pure Gaussian perturbation widely used in the literature to study the BH response~\cite{1970PhRvD...1.2870V,price:1972pw,MartelPoisson2005}. The parameter $r_0$ determines where the perturbation is placed relative to the potential barrier and  determines the functional form of the weight $W_n$ as discussed in Sec.~\ref{sec:r0}. 

\begin{figure}[!ht]
\centering
\includegraphics[width=\linewidth]{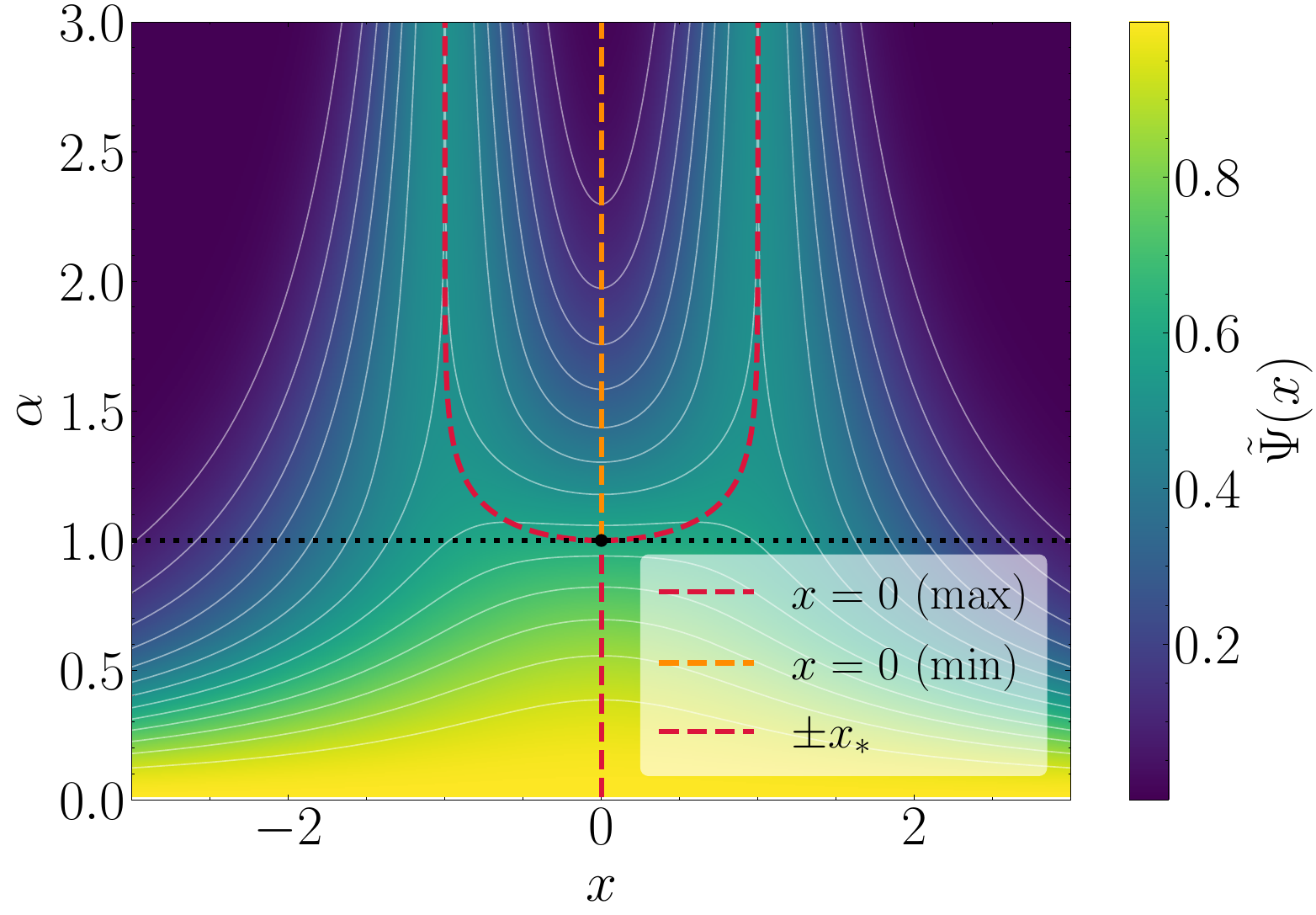}
\caption{\justifying 
Spectral structure of the ID $\tilde{\Psi}(x)$ as a function of $\alpha = \sigma\nu$ and $x = \omega/\nu$. Notice the transition at $\alpha = 1$ from a single maximum at $x = 0$ to two symmetric maxima at finite frequencies $x = \pm x_*$. The central extremum changes from a maximum for $\alpha < 1$ to a minimum for $\alpha > 1$, with $\alpha = 1$. This transition directly controls whether the BH response is tail-dominated ($\alpha < 1$) or QNM-dominated ($\alpha > 1$).
}
\label{fig:bifurcation_p1}
\end{figure}
A dimensionless parameter $\alpha \equiv \sigma\nu$ governs the spectral character of the ID: for $\alpha < 1$, the spectrum has a single maximum at $\omega = 0$ (broadband, low-frequency dominated); for $\alpha > 1$, two symmetric maxima appear near $\omega = \pm\nu$ (frequency selective, QNM-dominated). The parameter $\alpha$ counts the number of oscillation cycles contained within the Gaussian width $\sigma$. For a given $\sigma$, in $\alpha\ll1$ regime, the perturbation varies slowly across the envelope. Consequently, the spectrum is effectively non-oscillatory and is centered around $\omega=0$. As $\alpha$ increases, oscillations become more pronounced within the envelope. At $\alpha = 1$ within the width $\sigma$, exactly one oscillation cycle gets completed. In the regime $\alpha\gg1$, the source is strongly oscillatory, completing several cycles within the width.  As a result, its spectrum localizes away from $\omega=0$, around $\omega\simeq \pm\nu$.
In Fig.~\ref{fig:bifurcation_p1} we show this spectral structure of  $\tilde{\Psi}(x)$ in terms of $\alpha$ and the ratio $x\equiv\omega/\nu$. The color map determines the amplitude of $\tilde\Psi(\omega)$ for several combinations of $\alpha \in [0,3]$ and $x\in  [-3,3]$. The value $x=0$ provides a maximum in the amplitude for $\alpha<1$ and a minimum for $\alpha>1$. Notice that  two maxima appear at $x=\pm 1$ for $\alpha>1$ which disappear otherwise. $\alpha=1$ controls the balance between non-resonant BH tails and QNM-driven contributions to the BH response. 
Having established a class of perturbations with controllable spectral content, we now turn to the analytical evaluation of the excitation amplitude and make the spectral overlap interpretation explicit.

\section{Analytical Framework: Excitation as a Spectral selection process}
\label{sec:analytical}
\subsection{The overlap integral as a Fourier transform in the asymptotic limit}

In the asymptotic limit $r_\star \to \infty$, the QNM wavefunctions satisfy $\psi_{\rm in}(r_\star, \omega_n) \to A_{\rm out}(\omega_n) \, e^{i\omega_n r_\star}$, so the normalization cancels in Eq.~\eqref{eq:tn}, $W_n\to 1$ and the overlap reduces to standard Fourier transform
\begin{equation}
T_n = \int_{-\infty}^{+\infty}
      \mathcal{I}\big(\omega_n, r'_\star\big)\,e^{i\omega_n r'_\star}\,\d r'_\star\,.
\label{eq:tn_ft}
\end{equation}
The asymptotic approximation therefore neglects the near zone effects in the filtering process. This provides a useful approximation making the resonance conditions transparent and spectral structure evident.

A BH QNM at $\omega_n$ selectively samples the Fourier content of the source at that single complex frequency. The source spectrum acts as the input to the BH's bank of frequency-selective filters, one for each QNM, and $T_n$ is literally the output of the $n$-th filter. Designing the perturbation therefore means designing an input signal whose Fourier content matches the desired filter outputs. This is the language in which the results of this paper are most naturally understood.

For purely ingoing ID, the time derivative of the initial field is $\partial_t\Psi|_{t=0} = \partial_{r_\star}\Psi|_{t=0}$, and
a straightforward integration by parts (see Appendix~\ref{app:source})
shows that 
\begin{equation}
T_n = -2i\omega_n\int_{-\infty}^{+\infty}
      \Psi_0\big(r_\star\big)\,
      e^{i\omega_n r_\star}\,\d r_\star\,,
\label{eq:tn_ingoing}
\end{equation}
where $\Psi_0 = \Psi|_{t=0}$.
In deriving Eq.\eqref{eq:tn_ingoing}, an integration by parts has been performed, which in general produces a boundary term of the form $\left[ \Psi_0 e^{i\omega_n r_*} \right]_{-\infty}^{+\infty}$. For the localized Gaussian ID considered here, the rapid decay ensures that this term vanishes, and Eq.~\eqref{eq:tn_ingoing} is exact. For more general perturbations with slower asymptotic decay, this contribution may be non-zero and/or $\partial_t\Psi|_{t=0} = \partial_{r_\star}\Psi|_{t=0}$ may also not be applicable, which will be investigated in the future works.

Under the considered ID, the excitation amplitude is thus directly proportional to the Fourier transform of the initial profile in the asymptotic limit.
Note that this is a {\it spatial} Fourier transform of the ID evaluated at wave number $k=\omega_n$. In the asymptotic region $\omega=k$, so the spatial spectrum of the ID is the effective frequency domain input to the BH. Tuning $\nu$ is equivalent to tuning the frequency content delivered to each QNM pole.

\subsection{Analytic result for oscillatory Gaussian data}
\label{sec:analytic_prediction_resonance}

We now analytically evaluate this overlap explicitly for the oscillatory Gaussian ID introduced in Eq.~\eqref{eq:id}. 
Accounting for the symmetry of the QNM spectrum, $\omega_{n-} = -\omega_{n+}^*$, contributions from both frequency branches combine and give\footnote{For formal simplicity, we have ommitted here the term $e^{i \omega_n r_0}$, which adds a constant complex amplitude to all $C_n$'s computed in this work~\cite{Berti:2006wq,Andersson:1995zk}. The effects induced by a non-zero $r_0$ are discussed in Sec.~\ref{sec:results} .}
\begin{equation}
T_n = -i\sqrt{8\pi}\,\sigma\,\omega_n
      \bigg[e^{-\frac{\sigma^2}{2}(\omega_n + \nu)^2}
          + e^{-\frac{\sigma^2}{2}(\omega_n - \nu)^2}\bigg]\,.
\label{eq:tn_analytic}
\end{equation}
Near resonance, $\nu \approx \omega_n^{\rm Re}$, in the large $\alpha \gg  1$ limit, the excitation is dominated by the term involving $(\omega_n - \nu)^2$. The exact peak position depends on both the spectral width $\sigma$ and the imaginary part $\omega_n^{\rm Im}$, and is generally shifted slightly from $\omega_n^{\rm Re}$. The excitation is therefore maximized when the spectral support of the perturbation is concentrated near $\omega_n^{\rm Re}$.
Setting $\nu = 0$ recovers the pure
Gaussian result
\begin{equation}
\label{eq:tn_nu_zero}
\big|T_n\big|_{\nu=0} \propto \sigma\,\big|\omega_n\big|\,
e^{-\frac{\sigma^2 }{2}\mathcal{P}_n}
\end{equation}
where $\mathcal{P}_n \equiv
\big(\omega_n^{\rm Re}\big)^2 - \big(\omega_n^{\rm Im}\big)^2$. For weakly damped modes
($\mathcal{P}_n > 0$), the linear prefactor and the exponential
suppression compete, producing a maximum at
\begin{equation}
    \sigma^*_n = \frac{1}{\sqrt{\mathcal{P}_n}}\,.
\label{eq:optimal_width}
\end{equation}
This is the optimal pure-Gaussian width for exciting mode $n$: it is determined entirely by the mode’s complex frequency within the fixed normalization $A=1$. For strongly damped modes ($\mathcal{P}_n < 0$), the exponential factor enhances rather than suppresses the excitation, and $|T_n|$ grows monotonically with $\sigma$ without a maximum.

More generally, if the amplitude is allowed to scale with the width as $A \sim \sigma^p$, the excitation amplitude becomes
\begin{equation}
|T_n| \propto \sigma^{p+1} e^{-\sigma^2 \mathcal{P}_n/2},
\end{equation}
leading to an optimal width
\begin{equation}
\sigma^*_{n} = \sqrt{\frac{p+1}{\mathcal{P}_n}}.
\end{equation}
Therefore, while the Gaussian suppression factor is universal, the precise location of the optimal width depends on the normalization convention. In particular, the point-particle solution is recovered in the limit $p=-1$ and $\sigma \rightarrow 0$. In this work we restrict to the case $A=1$, for which Eq.~\eqref{eq:optimal_width} holds.

When $\nu \neq 0$, the near-resonance regime $\nu \approx \omega^{\rm Re}_n$ the second term exponentially dominates over the first exponential in Eq.~\eqref{eq:tn_analytic}. The factor $(\omega_n - \nu)^2$ in the exponent, evaluated at the complex QNM frequency, has real part $(\omega_n^{\rm Re} - \nu)^2 - (\omega_n^{\rm Im})^2$, which is minimized when $\nu = \omega_n^{\rm Re}$. This residual negative value means the exponent is never suppressed to zero at resonance,
preventing arbitrarily sharp enhancement. For finite $\sigma$, the exact maximum depends on $\sigma$ and is generally shifted from $\omega_n^{\rm Re}$. This shift decreases as $\sigma$ increases, and the peak approaches $\omega_n^{\rm Re}$ in the narrow-band limit.

\subsection{Beyond the asymptotic approximation: Leaver's wavefunctions}
\label{sec:leaver}

The analytic result of Eq.~\eqref{eq:tn_analytic}, derived under $W_n = 1$, requires the pulse to lie entirely in the far-field region. While large $\sigma$ narrows the spectral bandwidth and sharpens the resonance, the approximation requires the pulse to lie entirely in the far-field region. For large $\sigma$, the pulse extends over a wide spatial range, and its extended support produces a non-negligible overlap with the near-field region where $W_n \neq 1$, leading to systematic deviations between the asymptotic analytic prediction and both the Leaver computation and numerical results.

To assess the approximation's range of applicability, we also compute $C_n$ retaining the full radial structure of the QNM wavefunction via the Leaver series~\cite{leaver:1985ax,Leaver:1986gd}
\begin{equation}
C_n = B_n \int_{-\infty}^{+\infty}
      \mathcal{I}\big(\omega_n, r_\star\big)\,W_n\big(r_\star\big)\,
      e^{i\omega_n r_\star}\,\d r_\star\,.
\label{eq:cn_leaver}
\end{equation}
Details of the series construction are given in Appendix~\ref{app:leaver}. The Leaver computation is used as the primary theoretical reference, with the asymptotic formula serving as a transparent analytic approximation that exposes the physical mechanism. The comparison between the two is discussed in the results (Sec.~\ref{sec:resonance}) and provides a direct measure of near-field corrections to the spectral overlap picture.

The Leaver computation uses the exact radial eigenfunction for the QNM contribution through the weight $W_n$. However, the prompt response and late-time power-law tail, which arise respectively from high-frequency scattering and the branch cut in the complex frequency plane, are not represented by the QNM expansion and are absent from the Leaver reconstruction.
To validate these analytical predictions and assess their range of applicability, we now describe the numerical framework used to evolve the perturbations and extract the excitation coefficients.

\section{Numerical Framework}
\label{sec:numerical}
We solve Eq.~\eqref{eq:wave} numerically in the time domain by rewriting it as a first-order system and evolving it using a fourth-order Runge--Kutta (RK4) scheme. Spatial derivatives are computed with fourth-order finite-difference stencils on a uniform grid in the tortoise coordinate $r_\star$. The computational domain spans $r_\star \in [-200, 1500]\,M$ and the evolution is integrated up to a final time $T_F = 1500\,M$, which is sufficiently large to prevent contamination from boundary reflections during the time interval of interest. The effective potential is smoothly truncated near the boundaries by means of a window function, ensuring that it vanishes asymptotically and improving numerical stability. Unless stated otherwise, we extract the waveform at $r_\star^{\rm obs} = 100\,M$, which provides a reliable approximation to the asymptotic signal while avoiding near-field contamination. The dependence on the source position $r_0$ is tested in Sec.~\ref{sec:r0}.

The ID consist of a localized Gaussian wave packet. The conjugate momentum is set to $\Pi_0 = \partial_{r_\star}\Psi_0$, which corresponds to a purely ingoing wave packet in the flat-space limit. All numerical evolutions are performed with fixed amplitude $A=1$, matching the analytical normalization used throughout. In the presence of the effective potential the pulse is partially scattered, leading to a mixture of ingoing and outgoing components during the evolution. The time step is chosen according to a numerical Courant convergence condition, $\Delta t = \mathrm{CFL}\cdot\Delta r_\star$ with $\mathrm{CFL}=0.1$, ensuring stability of the evolution. No explicit boundary conditions are imposed: the large computational domain combined with the smooth truncation of the potential keeps spurious reflections at a negligible level within the time interval of interest.

The time-domain waveform exhibits three physically distinct contributions~\cite{Price:1971fb,Leaver:1986gd,Ching:1994bd,Andersson:1995zk,Andersson:1996cm}: an early-time prompt response, an intermediate QNM ringdown, and a late-time power-law tail. The prompt response is primarily determined by the direct propagation of the ID to the observer, with additional contributions from its early interaction with the effective potential, and arrives before the QNM-dominated regime is established. At late times, the power-law tail---sourced by backscattering off the long-range curvature potential---dominates and masks the exponentially decaying QNM contribution~\cite{Price:1971fb,Leaver:1986gd}. Between these two regimes the signal is well described by a superposition of damped sinusoids.

QNECs are extracted from the time-domain waveform by fitting Eq.~\eqref{eq:ringdown} at $r_\star=r^{\rm obs}_\star$,  using linear least squares over multiple overlapping time windows within the ringdown regime. Each window yields an estimate $C_n(t_0)$.
To reduce sensitivity to the choice of fitting interval, we repeat the fit over a large ensemble of windows and define the reported excitation coefficient as the median of $|C_n|$ over this ensemble. Uncertainty bands are obtained from percentile intervals of the resulting distribution, quantifying the systematic uncertainty associated with window selection.
The phase of each mode is obtained as $\arg(C_n)$, with uncertainties estimated from its variation across the same ensemble of fitting windows.

To analyze the phase of numerical $C_0$ we define 
\begin{equation}
|C_0|_w \;\equiv\; \big|\langle C_0\rangle_w\big|\,,\qquad\phi_w \;\equiv\; \arg\!\big\langle C_0\rangle_w\,,
\label{eq:phase:def_w}
\end{equation}
where window $w\in\mathcal{S}$ labels the family of fitting windows and $\langle C_0\rangle_w$ denotes the estimated $C_0$ in the window $w$. Further implementation details are provided in Appendix~\ref{app:extraction}.

\section{Results}
\label{sec:results}
\subsection{Waveform dependence on spectral parameters}
\label{sec:waveforms}

\begin{figure}[!ht]
\centering

\subfloat{%
\includegraphics[width=0.99\columnwidth]{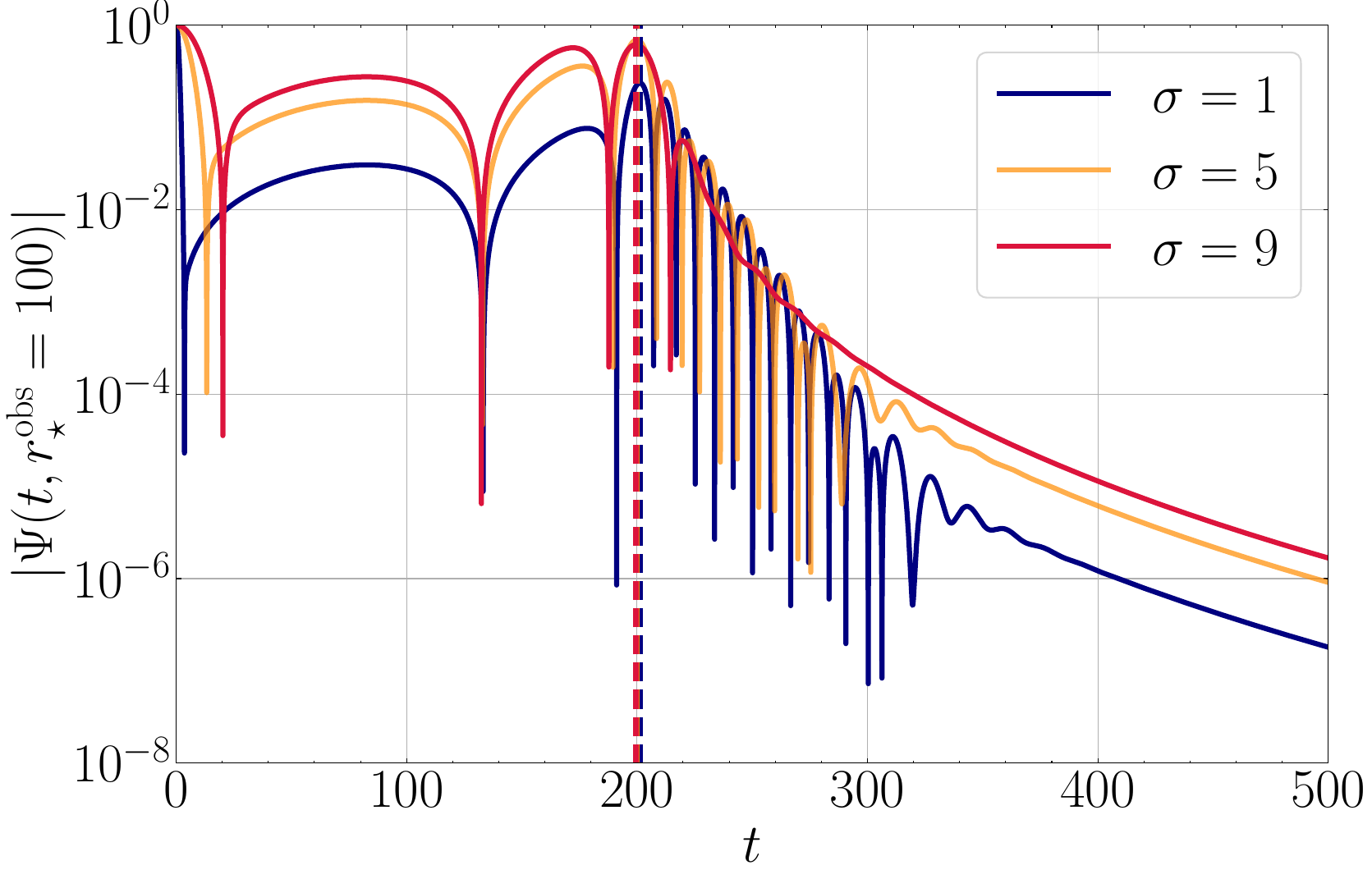}
}

\vspace{0.25cm}

\subfloat{%
\includegraphics[width=0.99\columnwidth]{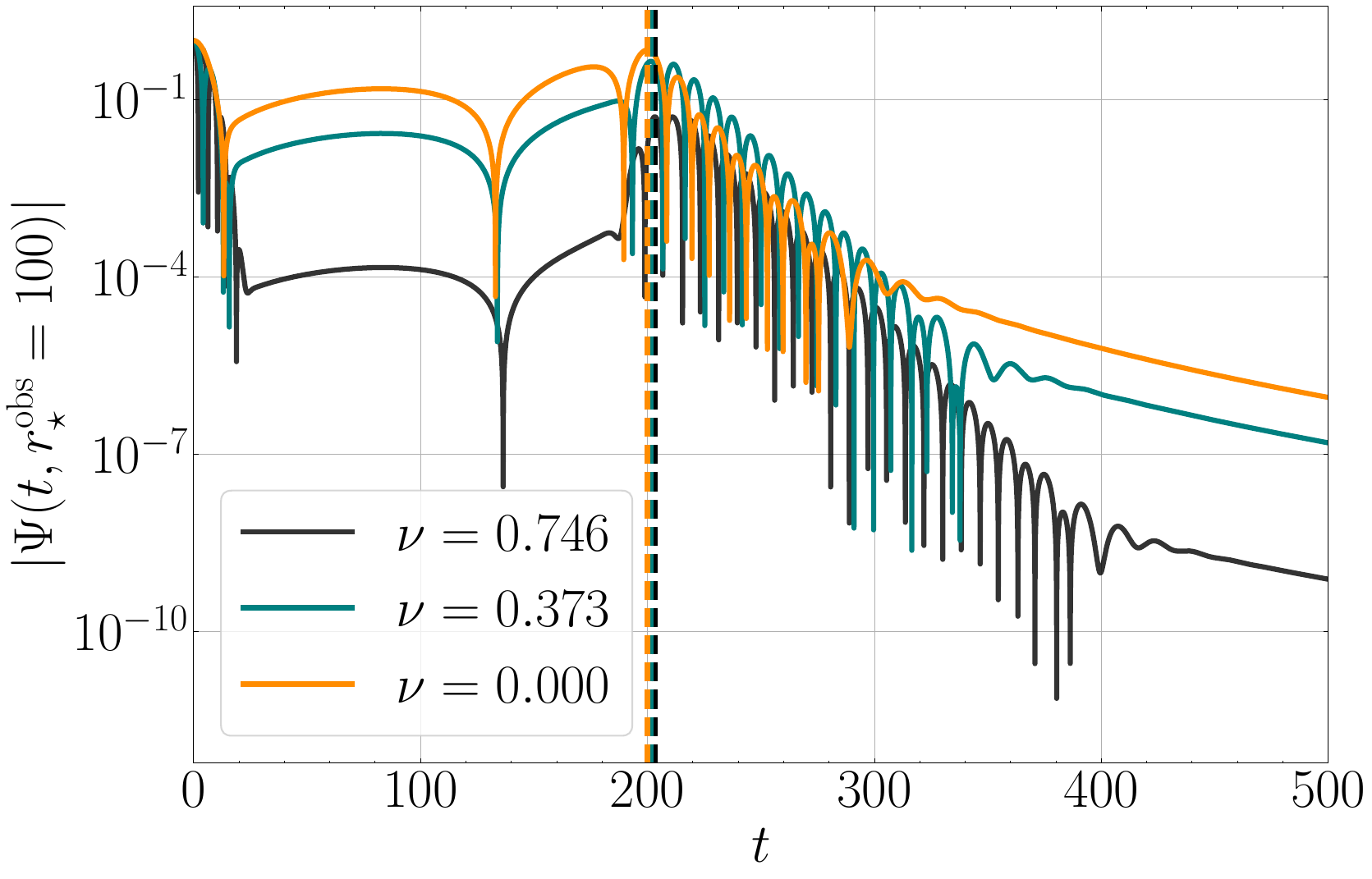}
}
\caption{\justifying
Waveforms extracted at $r_\star^{\rm obs} = 100M$ for the $\ell = 2$ mode.
\textbf{Top:} pure Gaussian ($\nu = 0$) for $\sigma = \{1,\,5,\,9\}$. Increasing $\sigma$ narrows the spectral bandwidth ($\Delta\omega \sim 1/\sigma$), suppressing QNM excitation and enhancing the late-time tail~\cite{tfm_alex}.
\textbf{Bottom:} oscillatory Gaussian with $\sigma = 5$ and varying $\nu$. Increasing $\nu$ shifts the spectral content toward finite frequencies ($\alpha = \sigma\nu$ increases from $0$ to $\sim 3.7$), suppressing the low-frequency tail and enhancing the QNM-dominated ringdown signal~\cite{tfm_alex}. The peak times identified within the ringdown window are indicated by vertical dashed lines, colored consistently with each corresponding waveform.}
\label{fig:waveforms}
\end{figure}

Figure~\ref{fig:waveforms} shows representative waveforms $|\Psi|$ extracted at fixed $r_\star^{\rm obs} = 100M$ for the $\ell = 2$ mode and $r_0 = 100$, illustrating the effect of varying $\sigma$ (top) and $\nu$ (down).
In the top panel ($\nu = 0$), the three curves correspond to $\sigma = \{1, 5, 9\}$. The quantity $t_{\rm peak}$ is the time at which the numerical maximum amplitude is observed. All three waveforms exhibit an initial prompt phase, peaking at $t_{\rm peak} \sim r_0 + r_\star^{\rm obs}$, where $|\Psi|$ reaches its maximum at $t_{\rm peak}\sim 200$, consistent with the propagation time of the perturbation to the light ring and back to the observer. Notice that the following trend emerges: increasing $\sigma$ narrows the spectral bandwidth and suppresses QNM excitation. For $\sigma = 9$, no sustained ringdown oscillations are observed; instead, the signal is dominated by a smooth, monotonic decay after the prompt peak.

This can be understood from Eq.~\eqref{eq:fourier_id}. At $\nu = 0$ ID's Fourier amplitude at $\omega = 0$ is $\Psi(0) \sim \sigma$, while the overlap amplitude $|T_n|$ decreases as $e^{-\sigma^2 \mathcal{P}_n/2}$ for large $\sigma$.  Conversely, for $\nu = \omega^{\rm Re}_n$ and $\alpha \gg 1$, Eq.~\eqref{eq:fourier_id} gives 
\begin{equation}
\label{eq:zero_freq}
\tilde{\Psi}(0) \propto e^{-\frac{\sigma^2}{2}\big(\omega^{\rm Re}_n\big)^2} \approx 0 \,,
\end{equation}
and the waveform is QNM-dominated with a negligible tail. Higher overtone QNECs can still grow in the large $\sigma$ regime. However, their rapid decay prevents them from contributing significantly to the waveform.

In the bottom panel ($\sigma = 5$), we simulate three initial pulses with $\nu =\{0,0.373,0.746 \}$, that are consistent with the zero, $\omega^{\rm Re}_{20}$ and $2\, \omega^{\rm Re}_{20}$  frequencies of the fundamental $\ell=2, n=0$ mode. As $\nu$ increases, the waveform transitions from a mixed, weakly oscillatory (orange and green curves) signal to a clean, QNM-dominated ringdown (black curve). Notice that, the QNM excitation amplitude of the green curve, with $\nu \approx \omega_0^{\rm Re}$, is enhanced with respect to the other black and orange curves, which follows from  Eq.~\eqref{eq:tn_analytic}. Moreover, the amplitude of the $\nu=0$ curve is higher than the amplitude for $\nu=0.746$ by a factor $\sim 2$.  Notice that, the exact ratio between the $\nu=0$ curve and a curve with $\nu=j\,\omega_0^{\rm Re}$ is
\begin{equation}
\frac{T_n\big(j\,\omega_n\big)}{T_n(0)}=\frac{e^{-\frac{\sigma^2\omega_n^2}{2}\big(j^2+2j\big)}+e^{-\frac{\sigma^2\omega_n^2}{2}\big(j^2-2j\big)}}{2}\,,
\end{equation}
where $j=0,1,2,3,\dots$ is a control parameter used to compare multiples of the fundamental frequency $\omega_0^{\rm Re}$. For $\sigma=5$ and $j=2$, the theoretical prediction gives $T_n(2\omega_n)/T_n(0) = 0.5$. We find that the numerical ratio of the medians reaches the value $|\Psi_{\nu=0.746}|/|\Psi_{\nu=0}| = 0.5$.
Similarly, the ringdown tails, emerging from the low-frequency branch-cut near $\omega=0$, are increasingly excited as $\nu \rightarrow 0$. 
For the $\nu=0$ case, Eq.~\eqref{eq:tn_analytic} predicts $T_n(\sigma=1)/T_n(\sigma=5)=1.068$. We find that the numerical waveform ratio reaches $|\Psi_{\sigma=1}|/|\Psi_{\sigma=5}| = 1.07$. For larger values of $\sigma$, the ringdown phase becomes extremely short-lived and rapidly transitions into a tail-dominated regime. In particular, the $\sigma=9$ waveform is already strongly contaminated by low-frequency branch-cut contributions shortly after its peak amplitude, making Eq.~\eqref{eq:tn_analytic} marginally/not applicable in practice for such broad ID.
\begin{figure}[!ht]
\centering
\includegraphics[width=\linewidth]{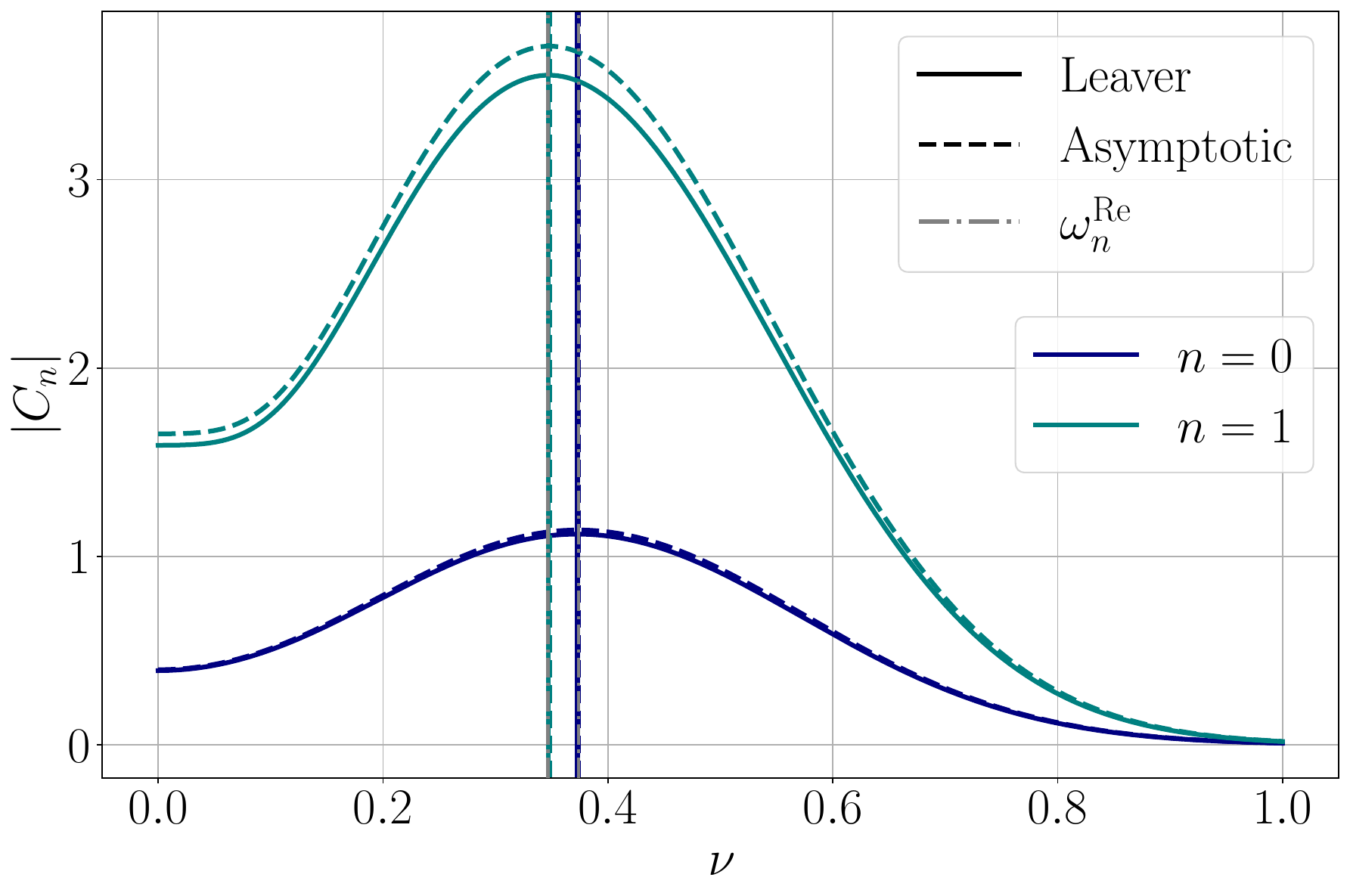}
\caption{\justifying
Quasinormal excitation coefficients $|C_n|$ as a function of the carrier frequency $\nu$ for oscillatory Gaussian ID with fixed width $\sigma = 5$ and $\ell = 2$. Vertical dotted lines indicate $\omega_n^{\rm Re}$ for $n = 0,1$. Each mode is maximally excited when $\nu \approx \omega_n^{\rm Re}$, demonstrating that the BH responds as a resonant spectral filter. }
\label{fig:qnecs_nu}
\end{figure}
\subsection{QNECs: the spectral filter in action}
\label{sec:resonance}

To quantify this behavior, we now turn from the time-domain waveforms to the corresponding QNECs, which provide a direct measure of the spectral filtering mechanism.
Fig.~\ref{fig:qnecs_nu} shows the amplitude of the QNECs $C_n$ for $\ell=2$ as a function of the excitation frequency $\nu$ and for $\sigma=5$. Results obtained using the Leaver and asymptotic methods are shown with solid and dashed lines, respectively, while the blue and green curves correspond to the modes $n=0$ and $n=1$. 


Each mode displays a clear resonance at $\nu \approx \omega_n^{\rm Re}$. The vertical blue and green dashed lines mark the known theoretical values for $\omega_{n=0,1}^{\rm Re}$ respectively. The asymptotic approximation reproduces the peak locations and overall structure for $\sigma = 5$, $r_0 = 100M$, showing an agreement with Leaver method to within $\sim 0.3\%$ in the resonance frequencies for $n=0$ and $n=1$, while the peak amplitudes agree within $\sim 2\%$ and $\sim 4\%$, respectively. Small deviations of $|C_n|$  near the peaks from the full Leaver computation, more pronounced for higher overtones, arise from near-field contributions to the QNM wavefunctions through the weight $W_n$, which become more relevant when $r_0$ is not very large (see Sec.~\ref{sec:r0}).

It is interesting to note that the amplitude of the tone $n$
grows as $n$ increases. In particular, near the  peak for $\alpha\gg1$
\begin{equation}
\big|A_n\big|_{\rm peak}
=
\sqrt{8\pi}\,\sigma |\omega_n|
\exp\Bigg(
\frac{\sigma^2}{2}\Big(\omega_n^{\rm Im}\Big)^2
\Bigg)\,,
\end{equation}
which increases with the overtone number $n$, as illustrated in the figure for $n=0,1$. The analytical QNECs contain a factor $e^{i\omega_n r_0}$, which has been absorbed into the initial time definition. For consistency the same scaling is applied to the Leaver and numerically extracted QNECs. Further details are provided in Appendix~\ref{app:extraction}.

\subsection{Resonance sharpening with bandwidth}
\label{sec:sharpening}

\begin{figure}[!ht]
\centering
\includegraphics[width=\linewidth]{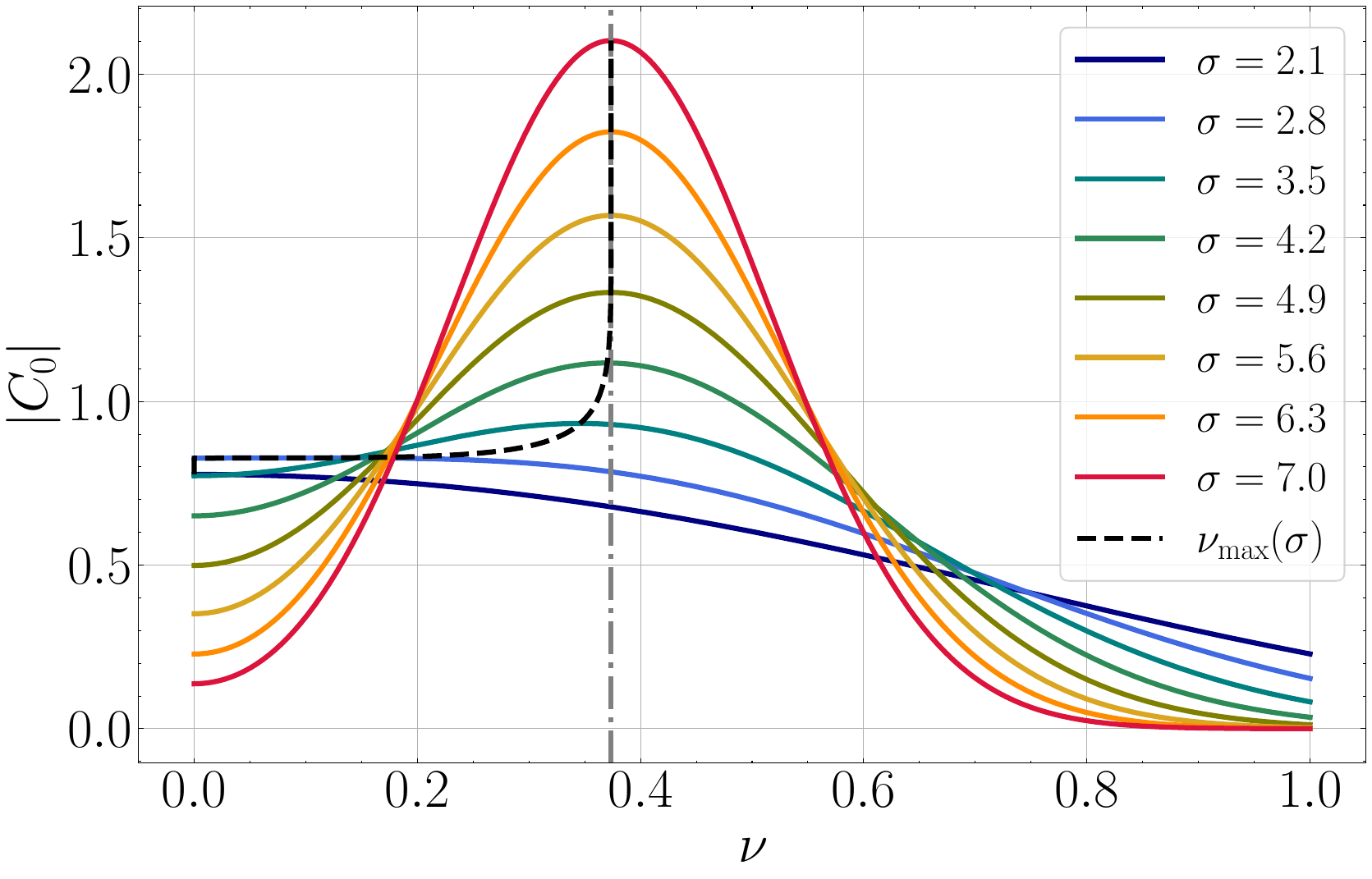}
\caption{\justifying
Fundamental-mode QNEC $|C_0|$ as a function of the carrier frequency $\nu$ for oscillatory Gaussian ID with varying width $\sigma$ (asymptotic approximation). As $\sigma$ increases, the spectral bandwidth narrows ($\Delta\omega \sim 1/\sigma$) and the resonance at $\nu = \omega_{20}^{\rm Re}$ (vertical dash-dotted line) sharpens. The dashed black curve shows the envelope of maximum excitation as $\sigma$ varies, highlighting how the optimal driving frequency approaches $\nu \simeq \omega_{20}^{\rm Re}$ as the bandwidth decreases.}
\label{fig:T0_nu}
\end{figure}

Fig.~\ref{fig:T0_nu} shows the fundamental-mode QNEC amplitude $|C_0|$ as a function of $\nu$ for a range of widths $\sigma \in [2,7]$ (solid lines).
In the small $\sigma$ regime (broad in the $\omega$ space), $T_0$ remains significant across a wide range of $\nu$ and the fundamental mode amplitude $C_0$ is maximally excited at $\nu< \omega_{20}^{\rm Re}$ even when the driving frequency is set to the QNM value (grey dash-dotted line). For $\sigma$ large (narrow  in the $\omega$ domain), the spectral bandwidth narrows ($\Delta\omega \sim 1/\sigma$), and the resonance sharpens around $\nu \approx \omega_0^{\rm Re}$. The excitation becomes increasingly selective: only perturbations whose carrier frequency aligns with the QNM frequency produce significant overlap, shown by the black curve asymptote, while contributions at other frequencies $\nu \neq \omega_{0}^{\rm Re} $ are exponentially suppressed by Eq.~\eqref{eq:zero_freq}. The amplitude of $C_0$ at its maximum value also increases with $\sigma$, while its width decreases significantly. 
Moreover, the $C_0$ response becomes more symmetric, exhibiting similar excitation strength at frequencies below and above the resonance $\omega_{20}^{\rm Re}$. In the neighborhood of the maximum
\begin{equation}
\big|T_n\big|
=\big|A_n\big|_{\rm peak}\,\,
e^{-\frac{\sigma^2}{2} \big(\nu-\omega^{\rm Re}_n\big)^2}\,,
\end{equation}
for $\alpha \gg1$. This symmetric response scales with $\sigma$ like a Gaussian of width $1/\sigma$. Therefore, as observed in the figure, sharpening the ID via increasing $\sigma$ allows the excitation profile to become increasingly localized around the resonant frequency $\nu \simeq \omega^{\rm Re}_{n}$. In the resonant regime with $\alpha \gg 1$, the finite width of each resonance peak is dominantly determined by $\sigma$, as observed in Fig.~\ref{fig:T0_nu}. 
The black dashed curve shows the location of the maxima, $\nu_{\max}(\sigma)$. For $\alpha<1$, the maximum occurs at $\nu_{\max}=0$. As $\sigma$ increases and $\alpha>1$, it shifts toward $\nu_{\max}=\omega_0^{\rm Re}$. Increasing $\sigma$ progressively narrows the carrier-frequency range $\Delta\nu$ that efficiently excites the QNM.

\subsection{Numerical validation: optimal width and resonance frequency}
\label{sec:validation}

The analytical framework of Sec.~\ref{sec:analytical} makes two sharp, parameter-free predictions. First, for a pure Gaussian ($\nu = 0$), the excitation amplitude is maximised at a width determined entirely by the complex QNM frequency, $\sigma^*_n = 1/\sqrt{\mathcal{P}_n}$ (Eq.~\eqref{eq:optimal_width}). Second, for an oscillatory Gaussian at fixed $\sigma$, the excitation peaks when the carrier frequency tracks the real part of the QNM frequency, $\nu \approx \omega^{\rm Re}_n$, with a residual shift induced by the imaginary part.
In this section we validate these predictions using numerical results.

\subsubsection{Spectral fingerprint of the QNM frequency.}

\begin{figure}[!ht]
\centering
\includegraphics[width=\columnwidth]{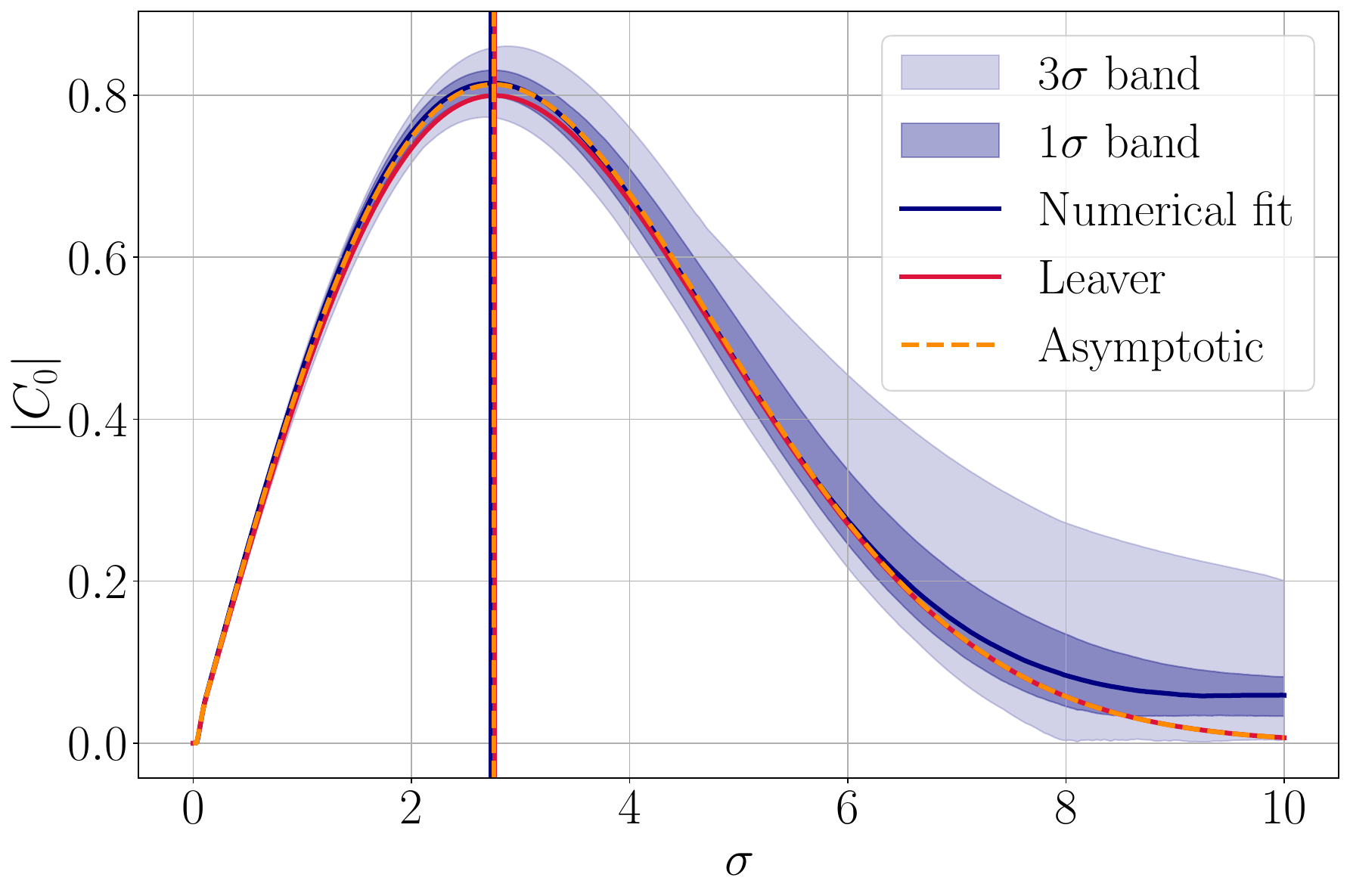}
\caption{\justifying 
Excitation amplitude of the fundamental mode $|C_0|$ as a function of the Gaussian width $\sigma$ for the pure Gaussian case ($\nu = 0$). The solid blue line shows the median of the numerical extraction over the family of fitting windows defined in Appendix~\ref{app:extraction}, the shaded blue regions indicate the $1\sigma$ and $3\sigma$ credible bands. The full Leaver prediction~\eqref{eq:cn_leaver} is shown as a solid red line, and the asymptotic approximation~\eqref{eq:tn_analytic} as a dashed orange line. Vertical lines mark the location of the maximum. }
\label{fig:C0_sigma}
\end{figure}

For the pure Gaussian case ($\nu = 0$) we vary the width $\sigma$ at fixed source location $r_0 = 100\,M$. 
Fig.~\ref{fig:C0_sigma} shows the resulting $|C_0(\sigma)|$ obtained by fitting the numerical extraction pipeline described in Appendix~\ref{app:extraction}. In particular, each waveform is first centred around its peak time $\bar t \equiv t-t_{\rm peak}$ and the QNM content is extracted through a sliding-window sequential fit of the form
\begin{equation}
\Psi\big(\bar{t}\,\big)=\mathrm{Re}\!\left[\sum_n C_n\,e^{-i\omega_n\bar t}\right].
\end{equation}
We extract $C_0$ by applying a sliding-window least-squares fit to each centred waveform. This procedure yields a set of estimates across different fitting windows, from which we construct a robust central value and associated uncertainty bands as described in Sec.\ref{sec:numerical}. The solid blue curve represents the median $|\widetilde{C}_0|$ over the window ensemble, taken as our central estimate. The solid red and orange dashed curves provide the Leaver and Asymptotic analytical estimates of $|\mathcal{C}_0|$. The shaded regions indicate the corresponding $1\sigma$ and $3\sigma$ credible bands (percentile intervals containing $68.27\%$ and $99.73\%$ of the empirical distribution), which quantify the systematic uncertainty associated with the choice of fitting windows. 

The numerical fit together with its uncertainty bands is broadly consistent with the Leaver and Asymptotic estimates for all the values of $\sigma$ considered. At $\sigma=0$, $|C_0|=0$ for all the curves considered (the zero perturbation limit). In the large $\sigma >> 1/\omega_{20}^{\rm R}$ regime, the Leaver and Asymptotic curves for $|C_0|$ decay exponentially following Eq.~\eqref{eq:tn_nu_zero}, while the Numerical fit saturates at a non-zero floor. This small deviation is sourced by the  extra power picked from the BH tail, which is genuinely present in the time-domain waveform but absent from both the Leaver and asymptotic constructions (see Fig.~\ref{fig:waveforms}). The width of the credible bands is also affected by the differences between low-sigma-high-sigma regimes. For $\sigma \lesssim 4$, where the waveform is QNM-dominated, the bands collapse onto the median and the extraction is essentially independent of the choice of fitting window. For $\sigma \gtrsim 5$, the bands widen progressively: the QNM identification becomes ambiguous because the ringdown is suppressed and the tail begins to drive the fit, and the spread of the bands is the empirical signature of this loss of identifiability.

\begin{figure}[!ht]
\centering
\includegraphics[width=\columnwidth]{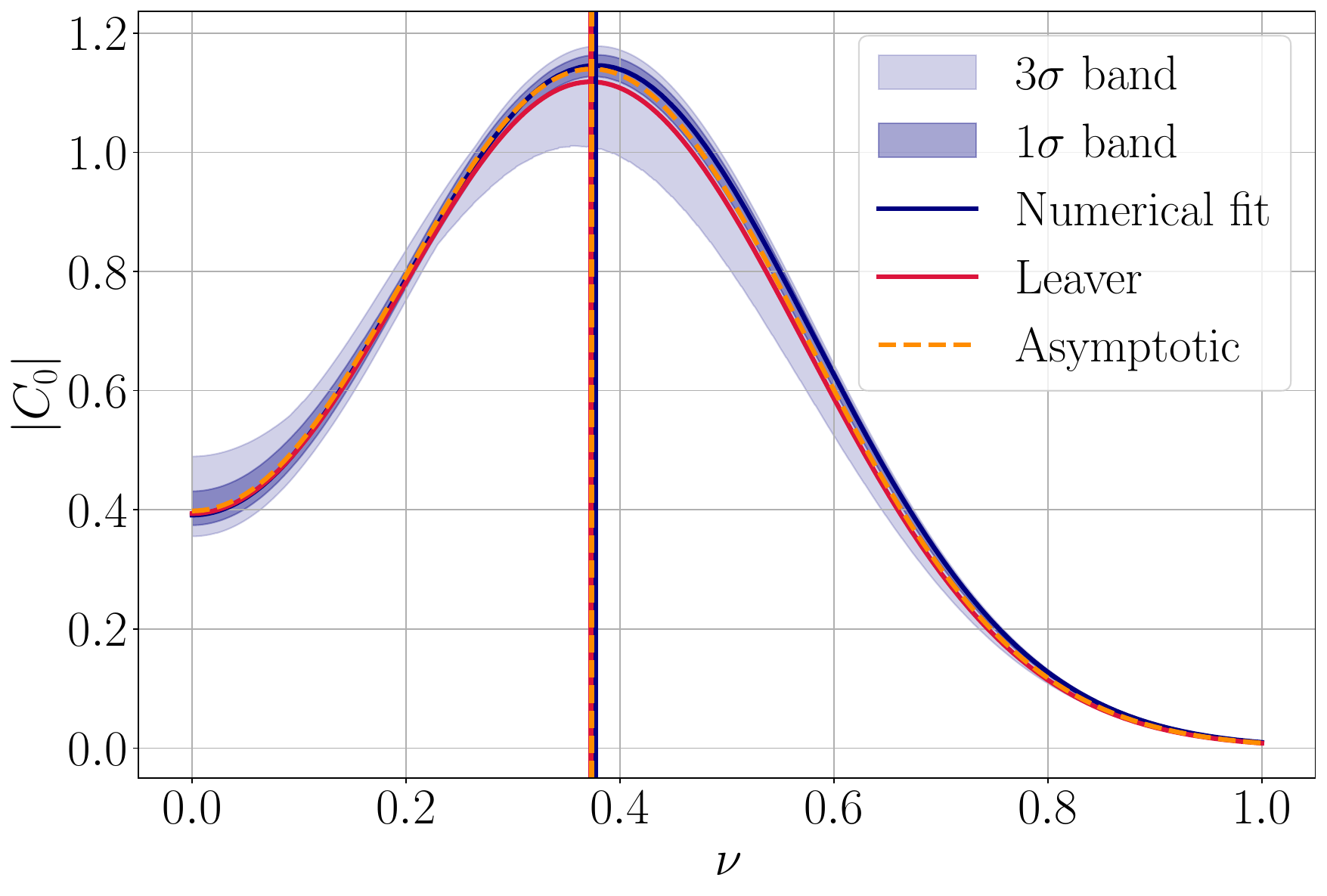}
\caption{\justifying 
Excitation amplitude of the fundamental mode $|C_0|$ as a function of the carrier frequency $\nu$ at fixed width $\sigma = 5$. The same analysis pipeline is applied as in Fig.~\ref{fig:C0_sigma}: the solid blue line shows the median, shaded regions indicate the $1\sigma$ and $3\sigma$ credible bands. The full Leaver prediction~\eqref{eq:cn_leaver} is shown as a solid red line, and the asymptotic approximation~\eqref{eq:tn_analytic} as a dashed orange line. Vertical lines mark the location of the maximum.}
\label{fig:C0_nu}
\end{figure}
The fit amplitude $|C_0|$ exhibits a well-defined maximum $|C_0|_{\rm max}$ that arises from the two opposing effects encoded in Eq.~\eqref{eq:tn_analytic}: the linear factor $\sigma|\omega_n|$ favours spatially wider pulses, while the Gaussian factor $\exp(-\sigma^2 \mathcal{P}_n / 2)$  suppresses the overlap once $\sigma$ exceeds $1/\sqrt{\mathcal{P}_n}$. Substituting the fundamental QNM frequency for $\ell = 2$ Schwarzschild, $\omega_0 \approx 0.374 - 0.089\,i$, gives $\mathcal{P}_0 \approx 0.132$. The asymptotic prediction yields a maximum at $\sigma^{*\,\mathrm{asy}}_0 = 2.755$~\cite{Andersson:1995zk}, while the full Leaver computation places it slightly higher at $\sigma^{*\,\mathrm{Leaver}}_0 = 2.762$. The small difference is primarily due to the inclusion of the near-zone weighting $W_n$. The numerical fit to the data locates the maximum at $\sigma^{*\,\mathrm{num}}_0 = 2.730$ thus, in agreement at the few-percent level with the analytical prediction.  We observe a small $\sigma_0^*$ shift on $|C_0|_{\rm max}$ of approximately $0.9\%$ between the numerical peak value and the corresponding Leaver and asymptotic predictions. 

The hierarchy of the three curves around the maximum is itself physically meaningful. The asymptotic approximation lies systematically above Leaver, with a relative excess of order a few percent near $\sigma^*_0$. This is the quantitative signature of the near-zone weighting function $W_n(r_\star)$: the asymptotic limit assumes $W_n = 1$, while Leaver retains the full radial dependence of the QNM wavefunction.
The departure of the dashed orange curve from the solid red curve is, in this sense, a direct outcome of the spatial structure of the QNM in the near zone (see Sec.~\ref{sec:leaver}). In the small-$\sigma$ limit the pulse is sufficiently narrow to lie entirely in the far field, $W_n \to 1$ effectively holds within its support, and the two analytical curves converge.
\subsubsection{Resonance frequency: spectroscopy of the BH}

We now fix $\sigma = 5$ and vary the driving frequency $\nu$, to scan the spectral content of the source across the QNM frequencies and its effects on the excitation amplitude $C_0(\nu)$. Fig.~\ref{fig:C0_nu} shows the resulting $|C_0|$ in terms of the excitation frequency $\nu$.

The three curves shown in Fig.~\ref{fig:C0_nu} (Numerical fit, Leaver prediction and asymptotic approximation) all exhibit the same qualitative shape: a single resonance peak around $\nu \approx \omega_{20}^{\rm Re}$, with a smooth decay in both sides. At low frequencies $\nu \to 0$ all three curves converge to the same value $|C_0|\approx 0.5$, while at high frequencies $\nu \to 1$ they decay together toward zero. The most visible difference among them is in the height of the peak: the asymptotic approximation (dashed orange) lies above Leaver (solid red), which in turn lies marginally above the numerical fit (solid navy). The credible bands remain narrow across the entire range of $\nu$, indicating that the extraction is robust everywhere.
The value of the Numerical, Leaver and Asymptotic curves near the peak mirrors the behavior observed in the Fig.~\ref{fig:C0_sigma}: the asymptotic approximation slightly overestimates the amplitude because it neglects the near-zone weighting $W_n$ compared to Leaver approach. The analytical and numerical curves are consistent to within the $1-\sigma$ bands for $\nu \in [0,1]$.
In $\nu\to0$ regime both terms in Eq.~\eqref{eq:tn_analytic} are equal and nonvanishing, while they exponentially decay in $\nu\to1$ regime. This happens as the ID's spectrum predominantly lies far from the QNM frequency.

Quantitatively, the three curves locate their maxima $\nu^{\rm asy}_{\rm max}=0.3736$, $\nu^{\rm Leaver}_{\rm max}=0.3727$, and $\nu^{\rm fit}_{\rm max}=0.3775$, all within $\sim 0.1\%$ of the real part of the fundamental QNM frequency $\omega_{20}^{\rm Re}=0.3736$. The agreement between the numerical extraction and the analytical predictions is at the percent level on the location of the maximum and on the overall lineshape.
The numerical fit peak does not sit exactly at $\omega_0^{\rm Re}$ but slightly below~\cite{Sieiro:2026:b-emris}, with a shift that vanishes only in the narrow-band limit $\sigma \to \infty$.
We observe that the credible bands across the entire $\nu$ range are narrower than these observed in Fig.~\ref{fig:C0_sigma}. At $\sigma = 5$ the pulse already lies well within the QNM-dominated regime ($\alpha = \sigma\nu \gtrsim 1$ for most of the $\nu$ range, see Fig.~\ref{fig:bifurcation_p1}), making the waveform QNM-dominated for all $\nu\gtrsim 0.2$. Thus $\nu$ acts as a clean spectral control at this $\sigma$.
\subsection{Phase analysis}
\label{sec:phases}
Being a complex number, along with the amplitude, the phase of the QNECs provides an additional probe of the spectral overlap mechanism.
The phase of each mode is obtained from the argument of the fitted coefficient $C_n$. Its uncertainty is estimated from the variation across the ensemble of fitting windows as discussed in Sec.~\ref{sec:numerical}.

\begin{figure}[!ht]
\centering
\includegraphics[width=\linewidth]{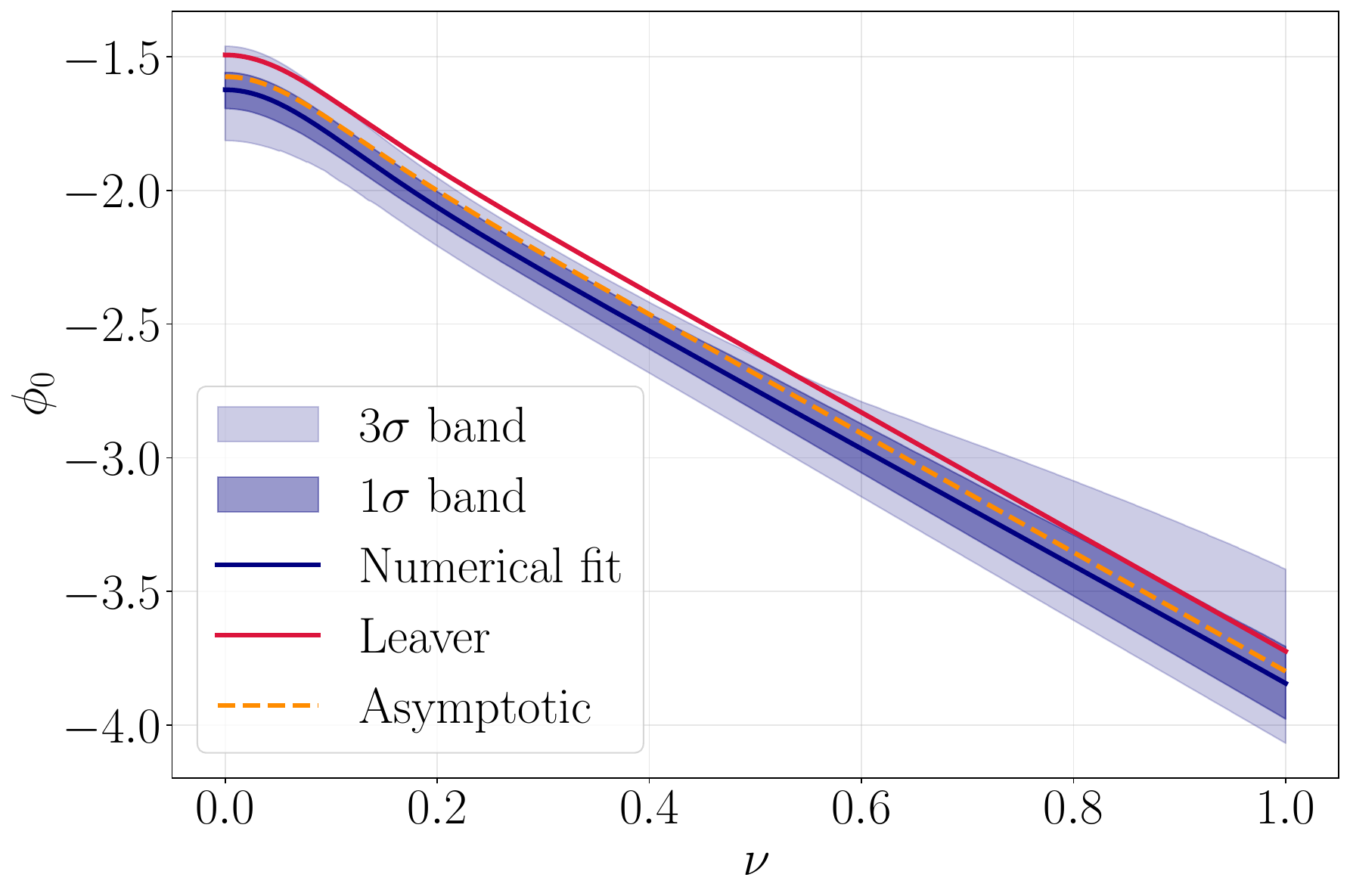}
\caption{\justifying
Fundamental-mode phase $\phi_0(\nu)$ for $\sigma = 5$ and $\ell = 2$. Solid blue: corrected numerical median, with $1\sigma$ and $3\sigma$ credible bands shaded. Solid red: full Leaver prediction. Dashed orange: asymptotic approximation.} 
\label{fig:phases}
\end{figure}

Fig.~\ref{fig:phases} shows the fit phase (blue color) with its corresponding error bands, compared to the phases obtained from asymptotic prediction (orange) and the full Leaver solution (red). The numerically fitted phase is fully consistent the two theoretical estimates used in this work for $\nu \in [0, 1]$. The phase decreases monotonically over the explored range of $\nu$, varying from $\sim -1.55$ to $\sim -3.78$ rad covering a range of $\sim2.2$ rad between $\nu=0$ and $\nu=1$. The residuals between the numerical phase and both analytical predictions remain below $\sim 6^\circ$ across the full interval.

The slope of the phase curve, i.e. $d\phi_0/d\nu$, remains approximately constant over the full range of $\nu$ considered, except $\nu\sim 0$. At $\nu\sim 0$ the phase becomes independent of $\nu$ approaching values of approximately $-1.60$, $-1.50$, and $-1.65$ rad for the asymptotic, Leaver, and numerical approaches, respectively. The asymptotic and the Leaver approach predicts an almost constant slope of phase $\sim -2.22$ and $\sim -2.20$ respectively. The median numerical value is $\sim -2.24$.
The degree-level agreement observed across the resonance region indicates that the analytical expressions of Sec.~\ref{sec:analytical} reproduce not only the magnitude of $C_0$, but also its phase structure.

A non-trivial feature of Fig.~\ref{fig:phases} is that the Leaver and asymptotic predictions are indistinguishable in the phase, even though they differ at the percent level in the amplitude near the maximum (Fig.~\ref{fig:C0_nu}). The two predictions differ by the QNM wavefunction factor $W_n(r_\star)$ inside the overlap integral (Eq.~\eqref{eq:cn_leaver}); 
$|C_0|$ remains sensitive to the detailed spatial structure encoded in $W_n$, whereas comparatively the phase $\phi_0$ appears to depend much more weakly on these near-zone corrections.

\subsection{Role of the source location}
\label{sec:r0}

\begin{figure}[!ht]
    \centering
    \includegraphics[width=\linewidth]{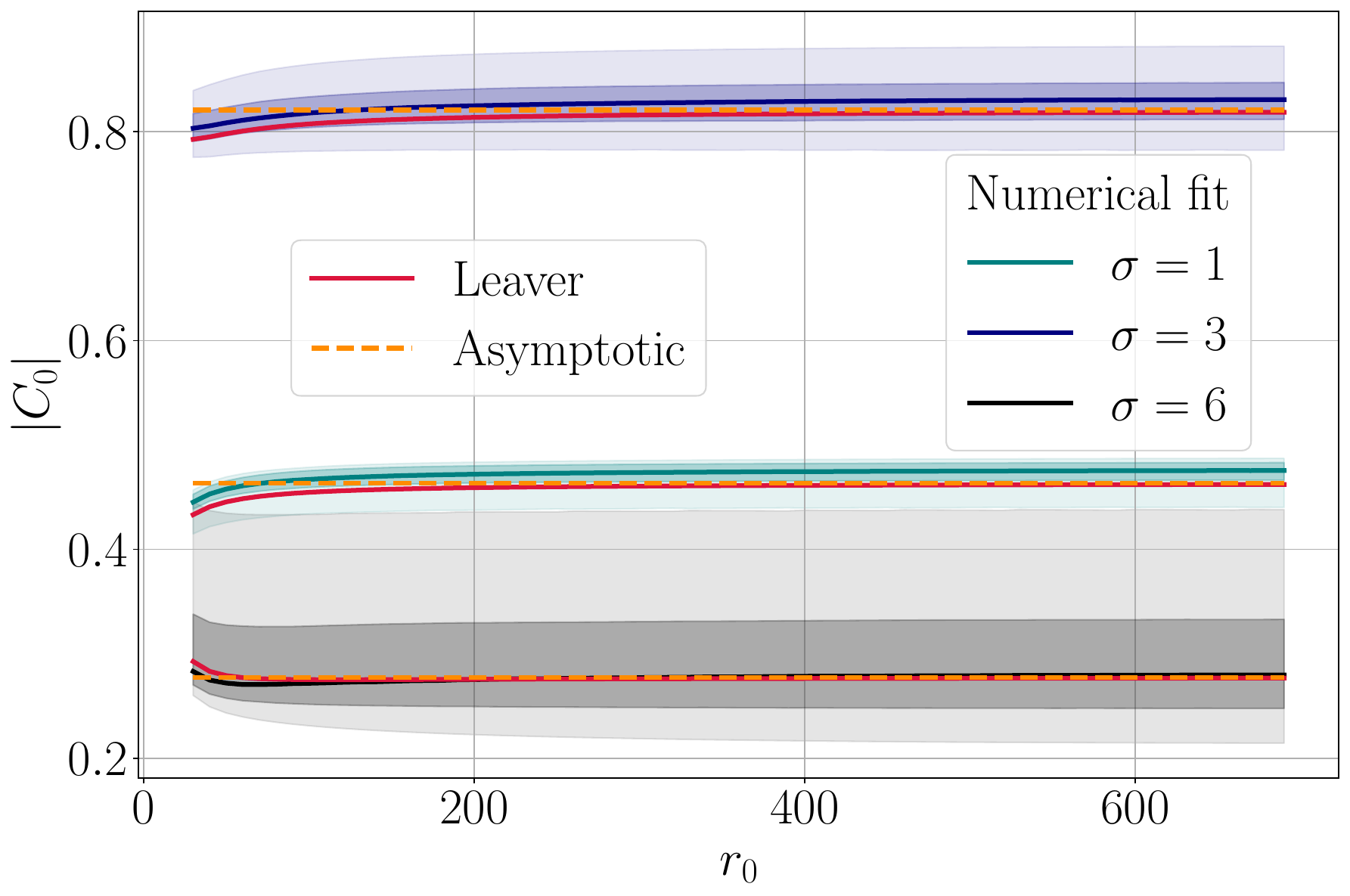}
    \caption{\justifying 
    Dependence of $|C_0|$ on $r_0$ for a pure Gaussian source ($\nu = 0$) at three widths $\sigma \in \{1, 3, 6\}$. Solid: numerical median, with $1\sigma$ and $3\sigma$ credible bands (same convention as Sec.~\ref{sec:validation}). Dashed: asymptotic prediction of Eq.~\eqref{eq:tn_analytic}. The asymptotic value is independent of $r_0$ by construction; its hierarchy can be deduced observing the resonance curve of Fig.~\ref{fig:C0_sigma}, with $\sigma = 3$ closest to the optimum $\sigma^*_0 \approx 2.76$. 
    }
    \label{fig:r0_dependence_sigma}
\end{figure}
Finally, we examine how the spatial location of the source affects the validity of the asymptotic approximation and the spectral filtering picture.
The asymptotic result of Eq.~\eqref{eq:tn_analytic} replaces the QNM weighting function by its far-field limit $W_n(r_\star) \to 1$, while the full Leaver computation of Eq.~\eqref{eq:cn_leaver} retains $W_n(r_\star)$. Both descriptions therefore hinge on the placement of the perturbation relative to the potential barrier at $r_\star^{\rm peak} \approx 3M$ and on the spatial extent $\sim \sigma$ of the pulse: the asymptotic approximation is well recovered whenever $\Psi(r_*)\approx 0$ near the tail of $V_\ell(r)$, while at small $r_0$ or large $\sigma$, part of the integral overlaps with the near zone where $W_n \neq 1$ and the asymptotic prediction breaks down systematically. To probe this picture quantitatively, we vary $r_0$ over a range spanning the transition between near- and far-field regimes and extract $|C_0|$ following the same pipeline used in Sec.~\ref{sec:validation}.

We consider two complementary configurations: (a) a pure Gaussian source ($\nu = 0$) at several widths $\sigma$, probing how the breakdown of the asymptotic approximation depends on the spatial extent and the position of the perturbation; and (b) an oscillatory Gaussian at fixed $\sigma = 5$ and several carrier frequencies $\nu$, testing whether the convergence to the asymptotic regime depends on the spectral content of the source.  Having different source location $r_0$ introduces large changes in the propagation time between the source and the observer. The fitting is therefore performed using a new time coordinate $\tau_{r_0}=t-r_\star^{\rm obs}-r_0$, that changes for each $r_0$ and sets appropriate time reference for each waveforms simulated (see App.~\ref{app:extraction}).

\begin{figure}[!ht]
    \centering
    \includegraphics[width=\linewidth]{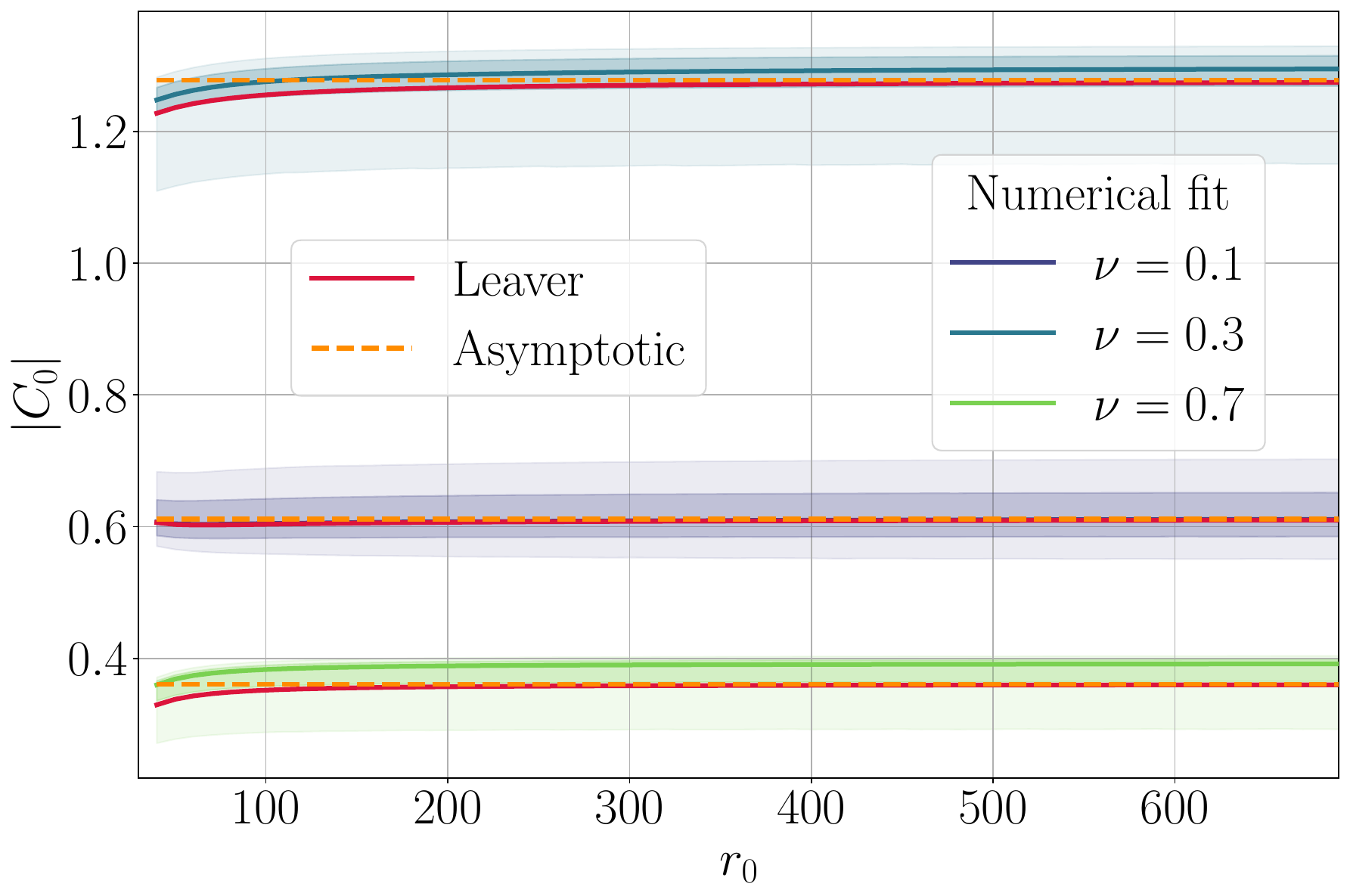}
    \caption{\justifying
    Dependence of $|C_0|$ on $r_0$ for an oscillatory Gaussian source at fixed $\sigma = 5$ and $\nu \in \{0.1, 0.3, 0.7\}$, spanning the resonance band of the fundamental mode at $\omega_0^{\rm Re} \simeq 0.374$. Solid curves show the numerical median, while the shaded regions correspond to the $1\sigma$ and $3\sigma$ credible bands, following the same convention as Fig.~\ref{fig:r0_dependence_sigma}. Dashed lines indicate the asymptotic prediction of Eq.~\eqref{eq:tn_analytic}. The asymptotic amplitudes follow the resonance curve of Fig.~\ref{fig:C0_nu}, with $\nu = 0.3$ closest to $\omega_0^{\rm Re}$ and therefore producing the strongest excitation. The low-frequency case $\nu = 0.1$ exhibits the largest deficit at small $r_0$, signalling the onset of the tail-dominated regime.  }
    \label{fig:r0_dependence_nu}
\end{figure}

Fig.~\ref{fig:r0_dependence_sigma} shows $|C_0|$ in terms of $r_0$ for three initial pulses with $\nu=0$ and $\sigma = [1,3,6]$. The purple, green and gray bands, show the numerical fits to $|C_0|$ while the solid red and orange dashed curves provide the Leaver and Asymptotic results. In particular, the opaque and light bands represent the $1\sigma$ and $3\sigma$ credible regions, respectively. We observe that all the curves remain approximately constant for $r_0 \gtrsim 90$. The asymptotic curve deviates slightly from the Leaver and numerical result around $r_0 \sim 90$,  due to the short-range effects induced by the nonvanishing tail of the RW gravitational potential at those radii. $\sigma = 3$ sits closest to the optimum $\sigma^*_0 \approx 2.76$ thus yielding the maximal QNEC amplitude among the three curves.
$\sigma = 1$ and $\sigma = 6$ sit on either side of the optimal excitation. The numerical curves converge toward the analytical results as $r_0$ increases, with the remaining differences in the median values lying within the uncertainty bands, which broaden for larger $\sigma$ due to tail contamination. 

Fig.~\ref{fig:r0_dependence_nu} shows the values obtained for $|C_0|$ as a function of $r_0$ for ID with $\sigma=5$ and $\nu = [0.1,0.3,0.7]$. We observe the same overall convergence toward the asymptotic limit as in Fig.~\ref{fig:r0_dependence_sigma}. The Leaver, asymptotic, and numerical results remain consistent at the $3\sigma$ level for all values of $r_0$ considered in this work, although the agreement becomes slightly weaker for $r_0 \lesssim 90$.  The curve with $\nu = 0.3$, corresponding to the value closest to $\omega_{20}^{\rm Re}$, produces the largest QNEC amplitude, followed by $\nu=0.1$ and $\nu=0.7$. The uncertainty bands remain approximately unchanged across the three cases considered, demonstrating the robustness of our fitting algorithm over the explored range of $\nu$.

\section{Multipolar hierarchy}
\label{sec:multipoles}
\begin{figure}[!ht]
    \centering
    \includegraphics[width=\linewidth]{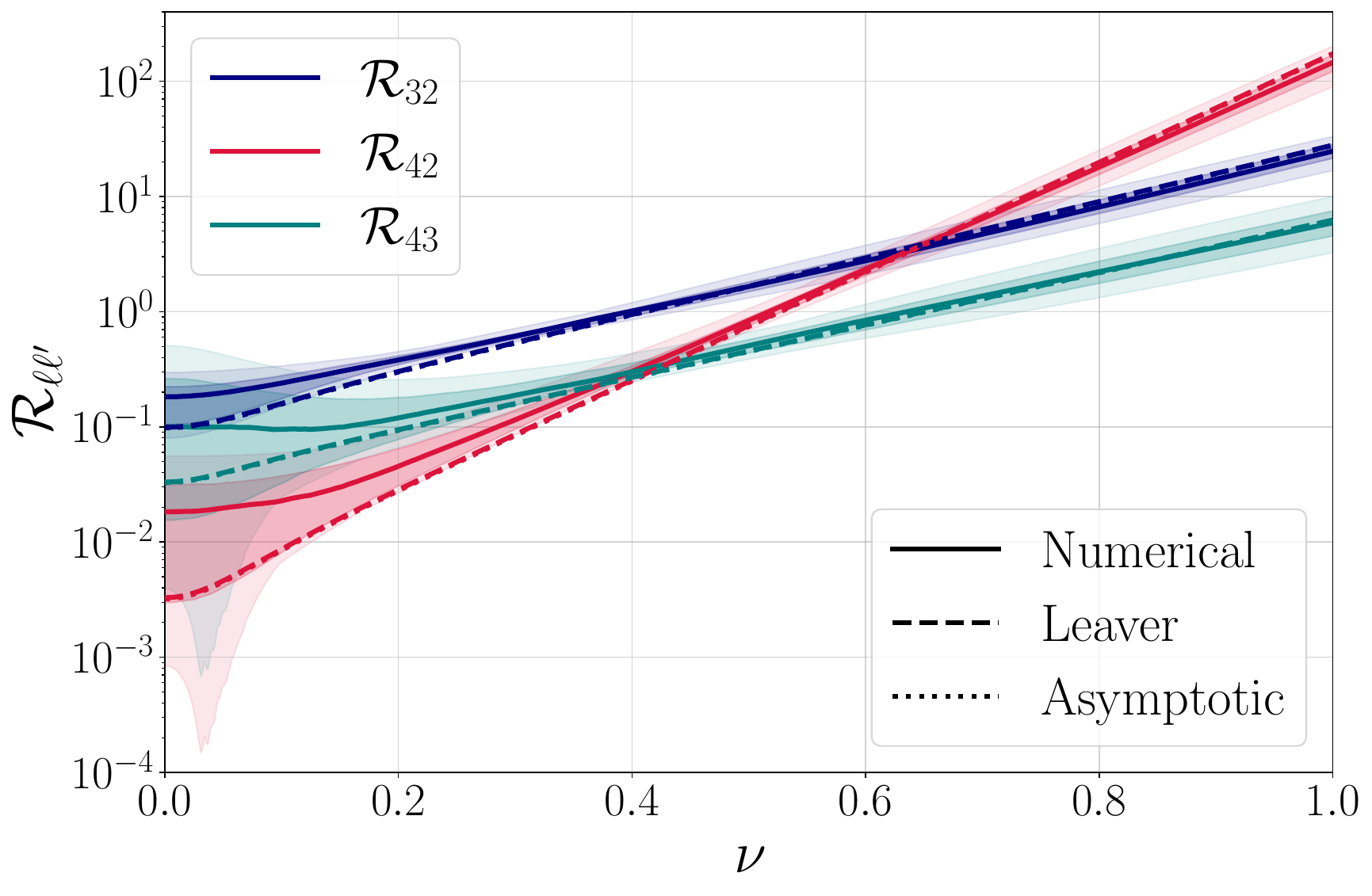}
    \caption{\justifying 
    Multipolar amplitude ratios $\mathcal{R}_{\ell\ell'}$ of Eq.~\eqref{eq:mult_ratios} as a function of $\nu$ at fixed $\sigma=5$, for the fundamental mode of each multipole. Solid lines: numerical median, with $1\sigma$ and $3\sigma$ credible bands. Dashed and dotted lines: Leaver and asymptotic predictions, respectively. The collapse of the three curves over four decades indicates that the near-zone weight $W_n^{(\ell)}$ largely cancels in the ratio, leaving the asymptotic approximation as a faithful description of the relative 
    multipolar excitation. }
\label{fig:qnec_ratios_multipoles}
\end{figure}
\subsection{Excitation ratios in higher multipoles}

In this section we extend the spectral filtering picture to higher multipoles, focusing on whether the same resonance mechanism applies independently to each angular sector. As a diagnostic we consider ratios of excitation amplitudes across multipoles
\begin{equation}
\mathcal{R}_{\ell \ell'} \equiv \frac{\left| C_{\ell} \right|}{\left| C_{\ell'} \right|}\,,
\label{eq:mult_ratios}
\end{equation}
which probe the relative excitation efficiency while reducing sensitivity to overall normalization. Fig.~\ref{fig:qnec_ratios_multipoles} shows the ratios $\mathcal{R}_{32}$, $\mathcal{R}_{42}$ and $\mathcal{R}_{43}$ as functions of $\nu$ at fixed $\sigma=5$, comparing numerical results with the full Leaver prediction and the asymptotic approximation.

The three ratios increase monotonically with $\nu$ and span more than four orders of magnitude across the explored range, reflecting the progressive excitation of higher multipoles as the spectral content of the source moves toward their respective resonance bands. At small $\nu$, the source is effectively a low-frequency Gaussian and overlaps preferentially with the lowest-$\ell$ QNM, suppressing all ratios. As $\nu$ approaches $\omega_{30}^{\rm Re}$ and $\omega_{40}^{\rm Re}$, the higher multipoles become preferentially excited and the ratios grow accordingly. The ordering $\mathcal{R}_{43} \ll \mathcal{R}_{32} \ll \mathcal{R}_{42}$ at large $\nu$ reflects the hierarchy of QNECs. Since $\omega_{40}^{\rm Re}>\omega_{30}^{\rm Re}>\omega_{20}^{\rm Re}$, larger $\nu$ favors larger $\ell$ QNMs to be excited with larger amplitude. Consequenctly $\mathcal{R}_{42}$ dominates in this range, followed by $\mathcal{R}_{32}$. Since $\ell=3,4$ QNMs are respectively $\sim 0.6$ and $\sim0.8$, both of them are sufficiently excited at higher $\nu$ range, thus reducing $\mathcal{R}_{43}$ comparatively. For the same rason $\mathcal{R}_{42}$ has the lowest value for $\nu\sim0$. In this neighborhood $\ell=2$ has the most spectral support. As all QNMs has the least spectral support in this range, $\mathcal{R}_{43} \sim \mathcal{R}_{32}$.

The agreement between numerical median, full Leaver prediction and asymptotic approximation is striking: the three curves coincide at the percent level across the full range of $\nu$. This confirms that the spectral filtering mechanism applies independently to each angular sector. More importantly, the indistinguishability of the Leaver and asymptotic predictions indicates that the near-zone correction $W_n^{(\ell)}$, which differs in detail between multipoles, largely cancels in the ratio: each $|C_\ell|$ is modified by its own near-zone weight, but the modifications are sufficiently similar across $\ell$ making the ratio insensitive to them.
The residual differences between numerical and analytical predictions is largely present only at small $\nu$ where the absolute amplitudes are smallest and also containing larger tail contamination discussed in Sec.~\ref{sec:r0}. 

\subsection{Amplitude--phase space}
\label{sec:amplitude_phase}
The two diagnostics developed in the previous subsections---the multipolar amplitude ratios $\mathcal{R}_{\ell\ell'}$ of Eq.~\eqref{eq:mult_ratios} and the QNM phases $\phi_{\ell}$ extracted in Sec.~\ref{sec:phases}---probe complementary aspects of the spectral filtering mechanism. The amplitude ratios encode the relative excitation efficiency of different angular sectors, while the phase combinations test the coherence of the excitation across multipoles. Taken together, they define a two-dimensional \emph{amplitude--phase space} that provides a more discriminating fingerprint of the ringdown than either observable in isolation.

To exploit this combined structure,  we define the multipolar phase combinations
\begin{equation}
\delta\phi_{\ell\ell'} \equiv \,\phi_{\ell} - \phi_{\ell'}\,,
\label{eq:phase_diffs}
\end{equation}
where $\phi_{\ell'}$ is the phase of the fundamental mode in the multipole ${\ell'=2}$, extracted with the procedure of Sec.~\ref{sec:phases}.

\begin{figure}[!ht]
    \centering
    \includegraphics[width=\linewidth]{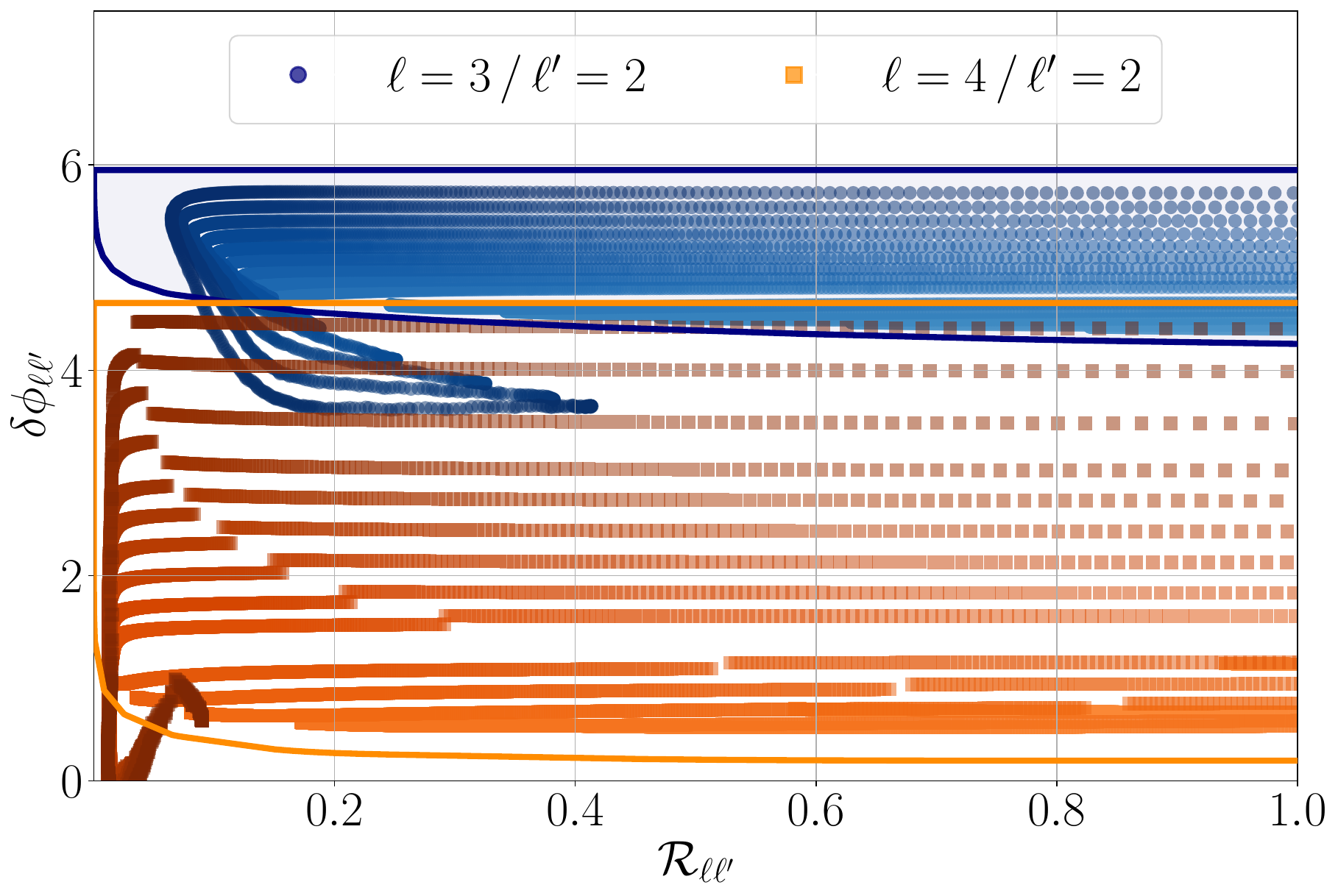}    \caption{\justifying  Amplitude-phase space structure of QNECs for $C_{0}$ and $\ell \in [2,4]$. Scatter distributions of $\ell=3/{\ell'}=2$ (blue) and $\ell=4/{\ell'}=2$ (orange) exhibit distinct clustering patterns in amplitude ratio versus phase difference across Gaussian ID profiles ($\sigma \in [1.00, 9.50]~M$). The shaded contours are instead constructed from the asymptotic QNEC predictions, providing the theoretical envelope associated with the numerical populations. Color gradient (dark to light) indicates increasing $\sigma$. Notice that $\delta\phi_{32}$ remains confined to a narrow band as observed in~\cite{Kubota:2025hjk,berti:2007fi} and in full NR binary simulations~\cite{Forteza:2022tgq,Cheung:2023vki}.}
\label{fig:amplitude-phase}
\end{figure}

Varying the carrier frequency $\nu$ and the width $\sigma$ of the oscillatory Gaussian source while keeping  $r_0$ fixed corresponds to scanning the spectral content of the perturbation across the resonance band, and at each $\nu-\sigma$ point, both $\mathcal{R}_{\ell \ell'}(\nu,\sigma)$ and $\delta\phi_{\ell \ell'}(\nu,\sigma)$ can be determined analytically by the asymptotic prediction~\eqref{eq:tn_analytic} and by the full Leaver computation~\eqref{eq:cn_leaver}. The result is a two-parameter band in the amplitude--phase plane, parameterised by $\nu$ and $\sigma$. In Fig.~\ref{fig:amplitude-phase} we show the $(\mathcal{R}_{\ell \ell'},\,\delta\phi_{\ell \ell'})$ plane for  $\nu \in [0,1]$ and $\sigma \in [1,9.5]$. For the modes $\ell=3$ and $\ell=2$, the plane covers only a small region of the $(\mathcal{R}_{\ell \ell'},\,\delta\phi_{\ell \ell'})$ parameter space, qualitatively similar to what is observed in realistic binary BH NR simulations and for the modes $\ell = m = 2$ and $\ell = m = 3$ ~\cite{Forteza:2022tgq,Cheung:2023vki}.

The shaded bands provide the same amplitude-phase space estimated with the asymptotic method and exhibit good agreement with the numerical results. Moreover, the range of values of $\nu-\sigma$ ($\alpha \in [0,9.5]$) used in our work span all the range of postmerger frequencies $\omega_m = m \Omega(t)$ observed in equal and near-equal mass realistic NR binary simulations~\cite{sxscatalog}. Therefore, this establishes a possible channel to infer information about the initial perturbation parameters $(\sigma,\nu)$ from the observed amplitude-phase space.
In contrast, the parameter space covered by $(\mathcal{R}_{42},\,\delta\phi_{42})$ is significantly broader, making it more difficult to infer the corresponding values of $(\sigma,\nu)$ from observations of $(\mathcal{R}_{42},\,\delta\phi_{42})$. These observations suggest that, for an appropriate choice of the $\ell$ modes, the filtering mechanism identified here may provide useful insight into the excitation of QNM's phase also in the highly nonlinear merger regime: the post-merger spacetime can be viewed as generating an effective perturbation whose spectral content drives a similar trajectory in $(\mathcal{R}_{\ell\ell'},\,\delta\phi_{\ell \ell'})$ space as our designed sources. We plan to explore these effects in a forthcoming work.

\section{Discussion and Conclusion}
\label{sec:conclusion}

We establish a unified picture of BH ringdown as a resonant spectral filtering process. The central result is that the overlap $T_n$ is a weighted spatial Fourier transform of the ID, evaluated at $k \sim \omega_n$, where the spacetime induces a mode-dependent weighting. The standard Fourier transform arises in the asymptotic limit. This makes the resonance condition transparent: spectral weight near $\omega^{\rm Re}_n$ enhances excitation, while weight away from $\omega^{\rm Re}_n$ suppresses it. 

The factorization $C_n = B_n T_n$ separates the intrinsic properties of the spacetime, encoded in $B_n$, from the spectral overlap with the perturbation, encoded in $T_n$. The parameters $\sigma$ and $\nu$ provide independent spectral control: $\sigma$ sets the bandwidth, while $\nu$ shifts the spectral power to enable resonant excitation of specific modes. 
We have studied the BH response coefficients $C_n$ across the $(\sigma,\nu)$ parameter space by: i) numerically solving the RW equation, and ii) using the analytic Leaver and asymptotic approximations. These predictions are confirmed numerically at the percent level: we find that all three methods remain consistent within the numerical uncertainty bands for the $C_n$ amplitude and phase. A similar agreement is observed when varying the source location, especially for large $r_0 \gtrsim 90$, where the source remains sufficiently localized and far from the tail of the RW potential.
Moreover, we have found that the optimal Gaussian width at which the resonance occurs, agrees with the analytic time-domain results to within $1\%$, and the resonance peak in $|C_0|$ as a function of $\nu$ is accurately reproduced by both the Asymptotic approximation and the full Leaver computation. We have also analyzed the higher-mode BH amplitude--phase parameter space. We find that, for a broad range of  $\nu$ and $\sigma$, only a narrow region of the amplitude--phase space of the $\ell=3,\,\ell=2$ plane is populated, exhibiting behavior qualitatively similar to that observed in binary BH NR simulations.

We have developed a new numerical algorithm, \texttt{QNMToolkit}, to perform robust fits to the numerical data. The algorithm automatically identifies the ringdown peak time and performs $\mathcal{O}(10^3)$ fits over an ensemble of sliding time-domain windows with variable sizes for each numerical waveform. The resulting collection of fits defines a statistical distribution for the extracted quantities, allowing us to agnostically characterize each fit parameter in terms of its median and variance. In this way, the method naturally accounts for uncertainties associated with the ambiguity in the choice of the fitting start time, contamination from prompt-response and tail contributions, and power leakage from higher overtones. The algorithm has been tested on time-domain waveforms obtained from numerical solutions of the RW equation. Nevertheless, we expect similar performance when applying it to the ringdown regime of more realistic NR waveforms, an extension that we plan to explore in future work. The algorithm is publicly available at~\cite{QNMToolkit2026}.

The spectral filtering picture provides a simple framework to relate the ringdown to the spectral content of the perturbation. The relevant quantity is the source spectrum evaluated at the QNM frequencies $\omega^{\rm Re}_n$: spectral weight near $\omega^{\rm Re}_n$ excites mode $n$, while weight near $\omega \sim 0$ feeds the late-time tail through the branch-cut contribution. Within this interpretation, different orbital configurations can be qualitatively associated with different spectral distributions. Quasi-circular inspirals generate radiation at frequencies $\omega_{\rm GW} \sim 2\Omega_{\rm orb}$ that approach the QNM band near merger, leading to significant overlap with $\omega^{\rm Re}_n$ and a QNM-dominated ringdown with a subdominant tail. In contrast, configurations with lower characteristic frequencies concentrate spectral weight at $\omega \ll \omega^{\rm Re}_n$, reducing the overlap with QNMs and enhancing the relative importance of the tail. These trends can be understood in terms of the parameter $\alpha = \sigma \nu$: perturbations with $\alpha \lesssim 1$ retain substantial low-frequency content and produce tail-dominated signals, while for $\alpha \gg 1$ the low-frequency component is suppressed, leading to cleaner QNM-dominated ringdown. 

Because the derivation relies only on the GF structure and the existence of QNM poles, the mechanism extends to other spacetimes, including rotating BHs. Thus, these results might have also direct implications for NR and GW data analysis. For a fixed total mass, narrow-bandwidth sources excite a limited set of modes, while broadband sources populate a wider overtone spectrum. The excitation and detectability of higher modes and higher overtones therefore depends on the presence of spectral power at their characteristic frequencies, which in turn depends on the mass ratio, spin, and orbital dynamics of the system. Extending this correspondence to realistic source models is a natural direction for future work.
More generally, the spectral filtering picture applies to any system governed by a wave equation with dissipative boundary conditions, with potential applications to nonlinear perturbations, mode coupling, and effective source modeling in systems such as extreme mass-ratio inspirals~\cite{Davis:1971gg,Taracchini:2014zpa,Thornburg:2019ukt,Nasipak:2019hxh,Lim:2019xrb, Apte:2019txp,DellaRocca:2025zbe,DeAmicis:2025xuh}. These aspects will be explored in the future.

\section*{Acknowledgements}
We thank Jo\~ao Sieiro dos Santos, Carlos Palenzuela Luque, Swetha Bhagwat and Lucía Vélez Tartajo for insightful comments on this work.
This work was supported by the Universitat de les Illes Balears (UIB) with funds from the Programa de Foment de la Recerca i la Innovaci\'o de la UIB 2024-2026 (supported by the yearly plan of the Tourist Stay Tax ITS2023-086); the Spanish Agencia Estatal de Investigaci\'on grants PID2022-138626NB-I00, RED2024-153978-E, RED2024-153735-E, funded by MICIU/AEI/10.13039/501100011033 and the ERDF/EU; and the Comunitat Aut\`onoma de les Illes Balears through the Conselleria d'Educaci\'o i Universitats with funds from the ERDF (SINCO2022/18146 - Plataforma HiTech-IAC3-BIO).
Project funded by the Comunitat Autonòma de les Illes Balears through a “Ajut per a projectes de recerca científica i tecnològica (PRD2025, PRD$2025\_099$)”.
X.J.\ and S.G\ are supported by the Spanish Ministerio de Ciencia, Innovaci\'on y Universidades (Beatriz Galindo, BG22-00034) and cofinanced by UIB.
S.D.\ acknowledges financial support from MUR, PNRR - Missione~4 - Componente~2 - Investimento~1.2 - finanziato dall'Unione europea - NextGenerationEU (cod.\ id.:\ SOE2024\_0000167, CUP: D13C25000660001).
The authors thankfully acknowledge the computer resources at MareNostrum~5 and the technical support provided by the Barcelona Supercomputing Center (BSC) through grants No.~RES-AECT-2025-1-0011, RES-AECT-2025-2-0038, and RES-AECT-2025-3-0050 from the Red Espa\~nola de Supercomputaci\'on (RES).
The authors acknowledge CINECA for providing high-performance computing resources and support through the ISCRA initiative under project HP10CU7X29.

\section*{Data availability}

There are no publicly available research data or software supporting this manuscript. Requests for further information or data should be sent to the authors.

\appendix
\section{Source overlap for purely ingoing initial data}
\label{app:source}
We show that for purely ingoing ID, the overlap integral reduces to twice the contribution from the initial field profile. The source is given by
\begin{equation}
\mathcal{I}\big(\omega, r_\star\big) = -i\omega\,\Psi_0 + \Pi_0\,,
\end{equation}
where $\Psi_0 = \Psi|_{t=0}$ and $\Pi_0 = \partial_t\Psi|_{t=0}$. For purely ingoing ID, imposing $\Pi_0 = \partial_{r_\star}\Psi_0$ and substituting into Eq.~\eqref{eq:tn_ft}, we obtain 
\begin{equation}
T_n = -i\omega_n \int e^{i\omega_n r_\star}\Psi_0\,\d r_\star
      + \int e^{i\omega_n r_\star}\partial_{r_\star}\Psi_0\,\d r_\star.
\end{equation}

Using
\begin{equation}
T_n^{(\Psi)} = -i\omega_n \int e^{i\omega_n r_\star}\Psi_0\,\d r_\star=-i\omega_n \sqrt{2\pi}A\sigma\tilde\Psi(\omega_n),
\end{equation}
and integrating the second term by parts we find
\begin{align}
\int e^{i\omega_n r_\star}\partial_{r_\star}\Psi_0\,\d r_\star = \Big[ &e^{i\omega_n r_\star}\Psi_0 \Big]_{-\infty}^{+\infty} +T_n^{(\Psi)}.
\end{align}
Assuming the pulse is localized so that $\Psi_0 \to 0$ as $r_\star \to \pm\infty$, the boundary term vanishes, yielding $T_n = 2 T_n^{(\Psi)}$. On the other hand, for purely outgoing ID, $\Pi_0 = -\partial_{r_\star}\Psi_0$, which leads to $T_n = 0$, confirming that outgoing pulses do not excite QNMs under the prescriptions assumed in this work.

\section{Computation of QNM wavefunctions via Leaver series}
\label{app:leaver}
The full QNM wavefunctions used in Sec.~\ref{sec:leaver} are computed following Leaver~\cite{leaver:1985ax,Leaver:1986gd}. The normalized wavefunction is 
\begin{align}
\psi_k\big(r_\star\big) = &\left(\frac{r-1}{r}\right)^{-2i\omega_k} \left(\sum_{j=0}^\infty a_j\right)^{-1} \sum_{j=0}^\infty a_j \left(1 - \frac{1}{r}\right)^j,
\label{eq:leaver_wf}
\end{align}
where the coefficients $a_n\left(\omega_k\right)$ satisfy the three-term recurrence
\begin{equation}
\alpha_n a_{n+1} + \beta_n a_n + \gamma_n a_{n-1} = 0,
\end{equation}
with
\begin{align}
\alpha_n &= n^2 + \big(2p+2\big)n + \big(2p+1\big)\,, \\[0.5em] \beta_n &= -\Big[2(n+2p)^2 + 2(n+2p) + \ell(\ell+1) - \varepsilon \Big]\,, \\[0.5em] \gamma_n &= n^2 + 4pn + 4p^2 - \varepsilon - 1\,,
\end{align}
where $p = -2iM\omega_k$ and $\varepsilon = s^2 - 1 = 3$ for gravitational perturbations. The series is evaluated via backward (Miller) recursion, normalized to $a_0 = 1$, and truncated at $N_{\rm max} = 300$, sufficient for convergence over the radial range and mode frequencies considered here. The normalization ensures $\psi_k \to 1$ as $r \to \infty$, consistent with Eq.~\eqref{eq:tn_ft}.

\section{Numerical extraction procedure for the QNM excitation coefficient}
\label{app:extraction}

For each simulation we find the time $t_{\rm peak}$ at which $|\Psi|$ reaches its maximum, for a  search window consistent with the propagation time $t_{\rm peak}\approx r_0 + r_\star^{\rm obs}$ from the ID to the observer. The time axis is then shifted to $\bar{t} \equiv t - t_{\rm peak}$, which approximately centres the onset of the ringdown at $\bar t\simeq0$.

We perform a sliding-window fit of a superposition of QNMs
\begin{equation}
\Psi\left(\bar{t}\, \right) \;\approx\; \mathrm{Re}\!\left[\,\sum_n C_n\, e^{-i\omega_n \bar{t}}\,\right],
\label{eq:app:fit_model}
\end{equation}
in a window $[t_0,\, t_0 + T_w]$ of fixed length $T_w = 100\,M$. The QNM frequencies $\omega_n$ are taken from tabulated values for the Schwarzschild background \cite{Stein:2019mop}. The window is slid over the waveform with a step $\Delta t_0 = 0.07\,M$, producing a discrete sequence of complex coefficients
\begin{equation}
\left\{ C_n\,\left(\bar{t}_0^{(i)}\right) \right\}_{i=1,\ldots,N_{\bar{t}_0}}, \qquad \bar{t}_0^{(i+1)} = \bar{t}_0^{(i)} + \Delta \bar{t}_0.
\label{eq:app:Ct0_sequence}
\end{equation}
A simultaneous fit of all QNM amplitudes is generically ill-conditioned because the QNM basis is strongly non-orthogonal over finite time intervals and neighbouring overtones possess similar damping rates. To reduce these degeneracies and for each fitting window, we first extract the fundamental mode $n=0$, subtract it from the waveform, and then fit the next overtone on the residual (which is a qualitatively similar procedure as described in~\cite{Ma_2022}). This procedure is iterated mode by mode up to the desired overtone order. 

\begin{figure}[!ht]
\centering
\includegraphics[width=1\linewidth]{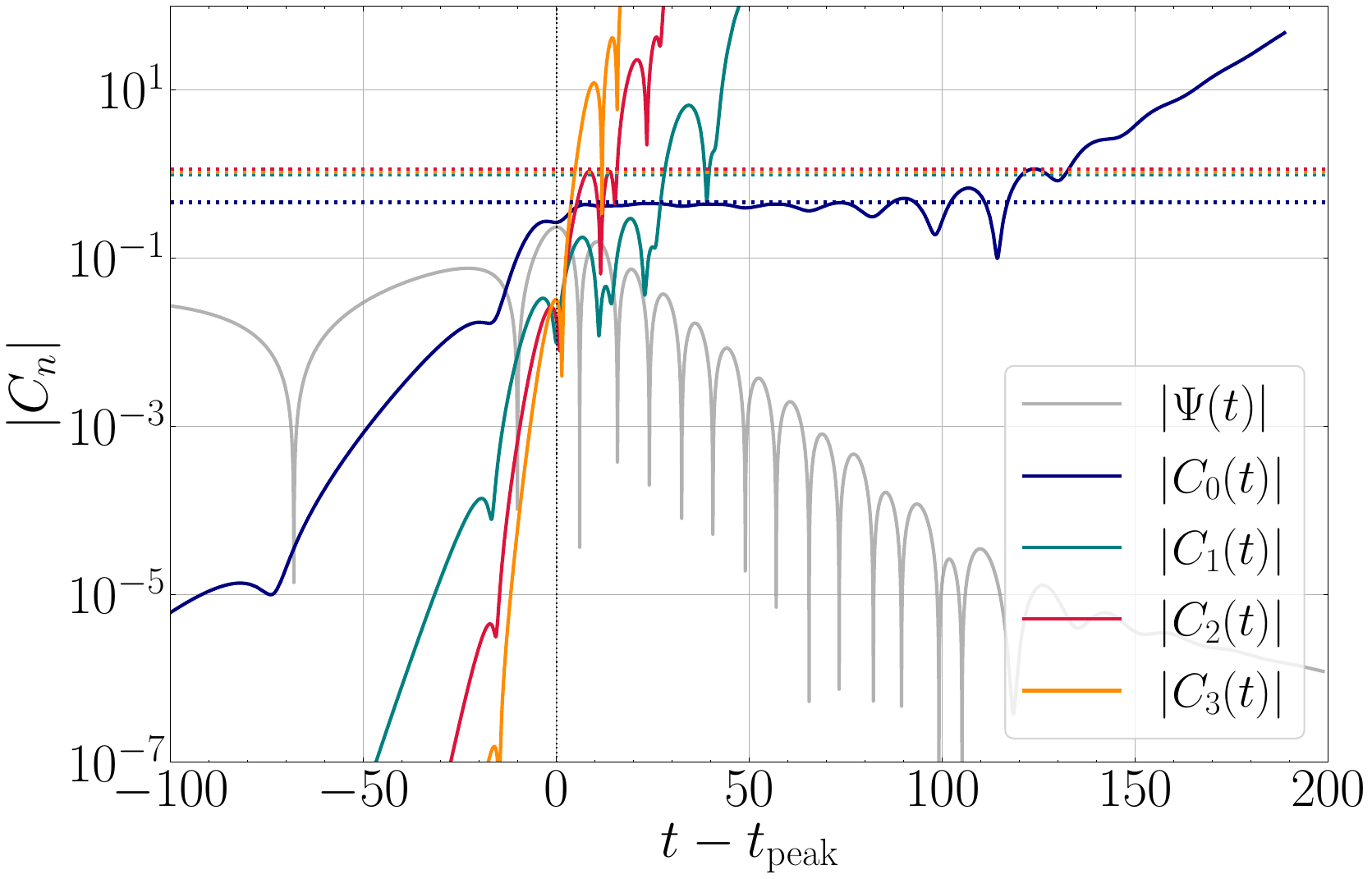}
\caption{\justifying 
Time evolution of the extracted QNM coefficients $|C_n(\bar{t}_0)|$ for $n = 0, 1, 2, 3$ (coloured curves), shown together with the underlying waveform magnitude $|\Psi(t)|$ (gray).}
\label{fig:Cn_t0_sequence}
\end{figure}

In Fig.~\ref{fig:Cn_t0_sequence}, we plot the extracted
$|C_n(\bar{t}_0)|$
for $n = 0, 1, 2, 3$ together with the underlying waveform $|\Psi(\bar{t})|$ as a function of $\bar{t}$. The fundamental coefficient
$|C_0(\bar{t}_0)|$
exhibits a clean plateau over an interval of order $\bar{t} \in [0, 60\,M]$, in which the extraction is essentially independent of $t_0$. At earlier times $\bar{t} \lesssim 0$, the fit is contaminated by the prompt response and
$|C_0(\bar{t}_0)|$ rises sharply because the fitting model attempts to absorb non-QNM prompt content into the fundamental mode.
At late times $\bar{t} \gtrsim 80\,M$, the underlying waveform transitions into the power-law tail and
$|C_0(\bar{t}_0)|$
drifts away from the plateau as the QNM identification becomes inaccurate. Higher overtones with $n \geq 1$ exhibit less stable behavior, making it difficult to identify a clear plateau region where
$|C_n| \approx \mathrm{constant}$. This qualitative structure --- prompt contamination at early times, a plateau-like QNM regime at intermediate times, and tail contamination at late times --- motivates the scan over multiple extraction windows described below.
 Similar instabilities have also been observed in full NR simulations, where possible sources of systematic error include multimode fitting degeneracies, waveform extraction effects, and ambiguities associated with the Bondi/BMS frame~\cite{Mourier:2020mwa,Forteza:2021wfq,Giesler:2024hcr,Mitman:2025hgy}.

\subsection{Statistical analysis over extraction windows}
\label{app:extraction:statistics}

A standard practice is to fix a single extraction window $[\bar{t}_0^{\min}, \bar{t}_0^{\max}]$ and quote a value of $|C_0|$ together with a statistical uncertainty derived from the dispersion of $C_0(\bar{t}_0)$ inside that window. This approach hides a strong sensitivity of the fit results to the choice of fitting window, thereby obscuring the systematic effects arising from variations of the window boundaries. To address this, we sample a large family of windows and treat the resulting set of extracted values as an empirical distribution. 

We define independent grids for the lower and upper window edges and introduce the set of valid pairs
\begin{equation}
\mathcal{S}
= \left\{ \left(\bar{t}_0^{\min}, \bar{t}_0^{\max}\right) \,\middle|\,
\begin{aligned}
&\bar{t}_0^{\min} \in \{0,\dots,45\} M, \\
&\bar{t}_0^{\max} \in \{15,\dots,75\} M, \\
&\bar{t}_0^{\max} > \bar{t}_0^{\min} \ge 20 M
\end{aligned}
\right\}.
\end{equation}
yielding $\mathcal{N}_w = 1485$ windows. We  impose a minimum duration on each  $\mathcal{S}$ of $\Delta T\equiv t_0^{\max}-t_0^{\min}\ge20\,M,$ which guarantees a sufficiently stable least-squares.

For each window $w\equiv(\bar{t}_0^{\min},\bar{t}_0^{\max})\in\mathcal{S}$ we compute the complex average of the $C_n$  coefficients provided in Eq.~\eqref{eq:app:Ct0_sequence}, which reads
\begin{equation}
\left\langle C_n\right\rangle_w\;\equiv\;\frac{1}{N_w}\sum_{\substack{i\,:\,\bar{t}_0^{(i)}\in[\bar{t}_0^{\min},\,\bar{t}_0^{\max}]}}C_n(\bar{t}_0^{(i)}),
\label{eq:app:estimator_complex}
\end{equation}
from which we obtain its modulus and phase,
\begin{equation}
|C_n|_w\equiv\left|\langle C_n\rangle_w\right|,\qquad\phi_w\equiv\arg\langle C_n\rangle_w,
\label{eq:app:estimator_modulus_phase}
\end{equation}
where $N_w$ denotes the number of samples contained in the window.
Since $C_n$ is complex, we  use of the complex average instead of directly averaging the amplitudes or phases. Moreover, not all windows in $\mathcal{S}$ lie within the clean ringdown region. The exact transition between the prompt-qnm and qnm-tail regimes is not well defined. We account for the contamination induced by the prompt response and the late-time tail by allowing our windows to partially overlap with these non-QNM regimes which provide us more realistic credible bands.

The collection $\{\langle C_0\rangle_w\}_{w\in\mathcal{S}}$ is then interpreted as an empirical distribution which median is defined as
\begin{equation}
\widetilde{C_n}=\mathrm{median}\Big(\langle C_n\rangle_w:w\in\mathcal{S}\Big),
\label{eq:app:median}
\end{equation}
while the $1\sigma$ and $3\sigma$ credible regions are obtained from the percentile intervals containing respectively $68.27\%$ and $99.73\%$ of the distribution. The resulting median curves are then compared directly with the analytical predictions discussed in the main text.

We center each value of the fit  phase around the median as
\begin{equation}
\delta_w \;=\; \arg\!\Big[\,e^{i(\phi_w - \widetilde{\phi})}\,\Big]\in (-\pi,\pi]\,,
\label{eq:app:phase:residual}
\end{equation}
with $\tilde{\phi}=\rm{arg}[\tilde{C}_n]$ to remove the $2\pi$ defegeneracy and obtain a branch centred around $\widetilde{\phi}$. Percentiles of the empirical distribution $\{\delta_w\}$ are then computed and mapped back to the absolute phase scale through $\widetilde{\phi}+\delta_w$, yielding the corresponding $1\sigma$ and $3\sigma$ credible intervals.
Alternative fitting strategies based on variable projection and nonlinear multimode optimization have also been explored in the recent ringdown literature~\cite{varpro}.
\subsection{Matching theoretical and numerical $C_n$}
\label{sec:pipeline:alignment}
The fitting described above is performed around the peak-centred coordinate $\bar t = t-t_{\rm peak}$ to consistently compare numerical extractions obtained from different waveforms  with different parameters $(\nu,\sigma)$. The position of this peak for each method used to solve the RW equation is consistent with $t_{\rm peak}^{\rm m}=r_\star^{\rm obs}+r_0+\delta t^{\rm m}$ with $\rm m$ standing for numerical, Leaver, and asymptotic, i.e., approximately the time required for the perturbation to propagate and interact with the peak of the potential barrier located near $3M$. Here, $\delta t^{\rm m}$ denotes a small method-dependent temporal offset, which generally differs among the approaches considered in this work and which needs to be accounted for to match the theoretical and numerical results.  Otherwise, even small misalignments in $t_{\rm peak}$ lead to exponentially amplified deviations between the theoretical and numerical $C_n$. Therefore, to consistently compare the different methods, we apply the following additional correction
\begin{equation}
    C_n^{\rm m}=C_n\, e^{i \omega_n \delta t^{\rm m}}\,,
\end{equation}
thereby mapping the theoretical and numerical waveforms onto a common temporal reference frame.

\section{Extension to higher multipoles}
\label{app:multipoles}

The analysis of Sec.~\ref{sec:validation} was carried out exclusively for the dominant quadrupolar sector $\ell = 2$, where the QNM excitation is largest and the time-domain extraction is less affected by the numerical noise. The simulations have been executed using the same parameters as for $\ell=2$, i.e. $r_0=100$ for the initial location of the pulse and $r_\star^{\rm obs}=100$ for the waveform extraction. In \ref{sec:multipoles} we also compared relative amplitude and phases for different multipole, i.e. $\ell=3,4$. For higher multipoles, we adopt mode-dependent extraction windows to balance the need to resolve the fundamental QNM overtone against numerical truncation errors: $T_{\max} = 75\,M$ for $\ell=2$, $T_{\max} = 40\,M$ for $\ell=3$, and $T_{\max} = 50\,M$ for $\ell=4$.
In this appendix we provide amplitude and phase for individual multipoles in a simialr manner to Sec.~\ref{sec:validation}.

\begin{figure}[!ht]
\centering
\subfloat{%
\includegraphics[width=\linewidth]{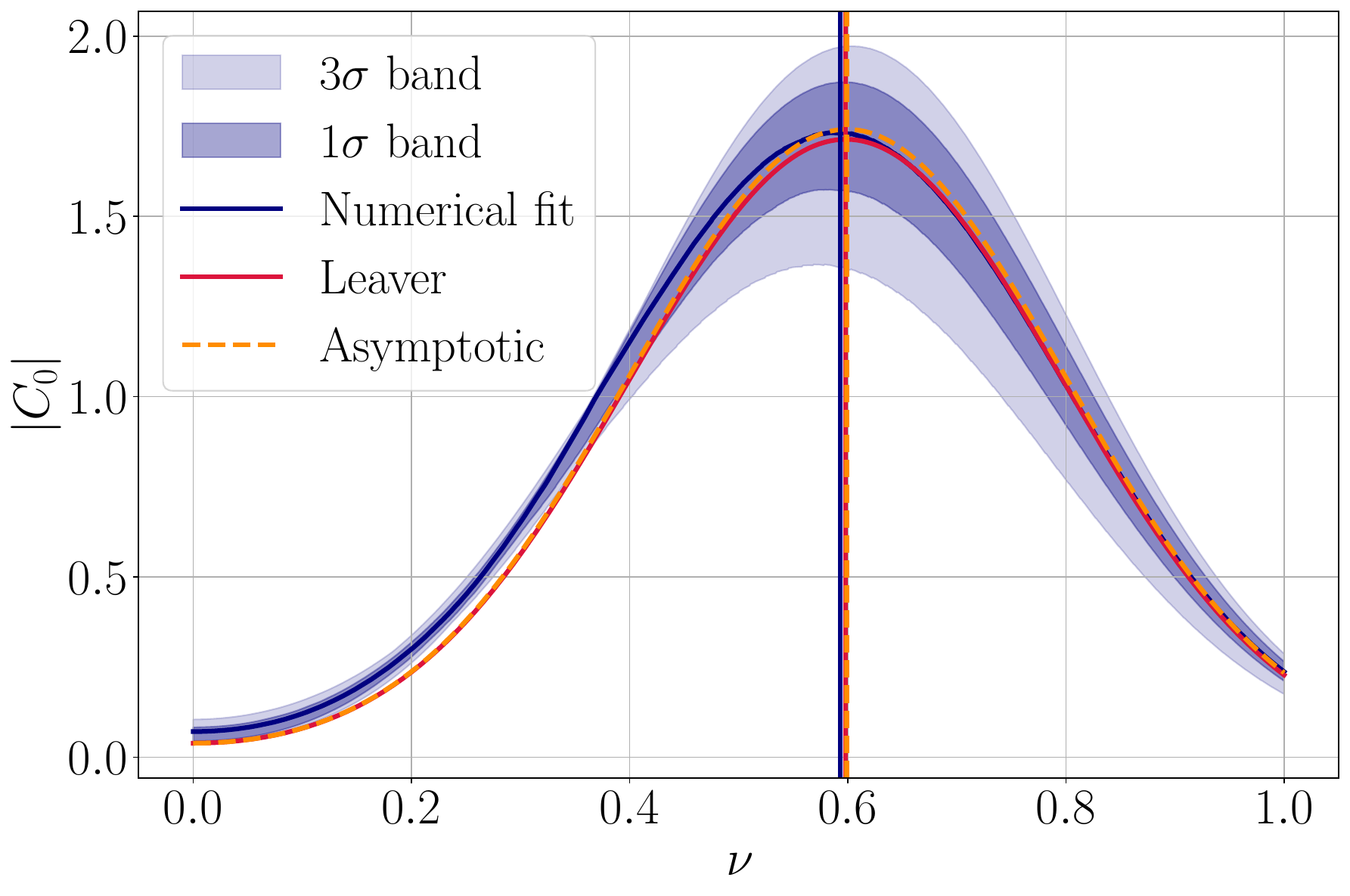}
}
\vspace{0.1cm}
\subfloat{%
\includegraphics[width=\linewidth]{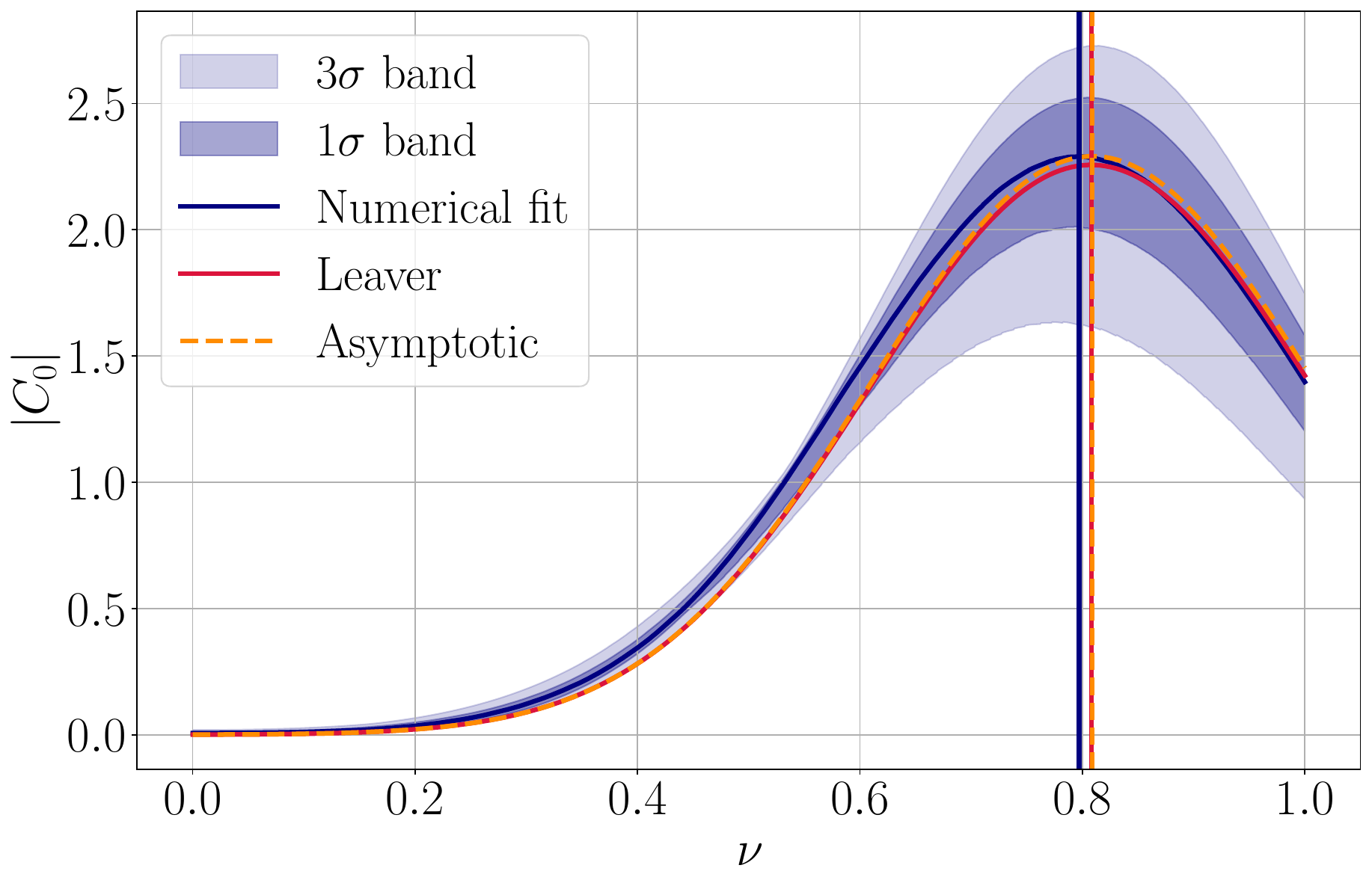}
}
\caption{\justifying 
Excitation amplitude of the fundamental mode $|C_0|$ as a function of the carrier frequency $\nu$ at fixed width $\sigma = 5$ for the higher multipoles \textbf{(top)} $\ell = 3$ and \textbf{(bottom)} $\ell = 4$. The conventions follow those of Fig.~\ref{fig:C0_nu}: solid blue is the numerical median with $1\sigma$ and $3\sigma$ credible bands, the solid red line is the full Leaver prediction~\eqref{eq:cn_leaver}, and the dashed orange line is the asymptotic approximation~\eqref{eq:tn_analytic}. The resonance peak shifts to higher $\nu$ as $\ell$ increases, in correspondence with the higher real parts of the fundamental QNM frequencies, $\omega_{30}^{\rm Re} \simeq 0.599$ and $\omega_{40}^{\rm Re} \simeq 0.809$.}
\label{fig:C0_l3_l4}
\end{figure}

Both panels of Fig.~\ref{fig:C0_l3_l4} exhibit the same qualitative structure as the quadrupolar case: a well-defined resonance peak whose position is set by the real part of the fundamental QNM frequency of the corresponding multipole, with the asymptotic and Leaver predictions tracking the numerical median at the percent level across the full range of $\nu$. The peak shifts to higher $\nu$ as $\ell$ increases, in agreement with the well-known monotonic growth of $\omega_0^{\rm Re}$ with $\ell$. The hierarchy of the three curves is also preserved: the asymptotic approximation lies systematically above Leaver near the maximum, with a relative excess at the percent level that quantifies the near-zone weighting $W_n^{(\ell)}$ for each multipole. The agreement is non-trivial: the reproduction of the numerical $|C_0(\nu)|$ across two additional multipoles, with peak positions, peak amplitudes and overall lineshapes all captured at the percent level.

\begin{figure}[!ht]
\centering
\subfloat{%
\includegraphics[width=\linewidth]{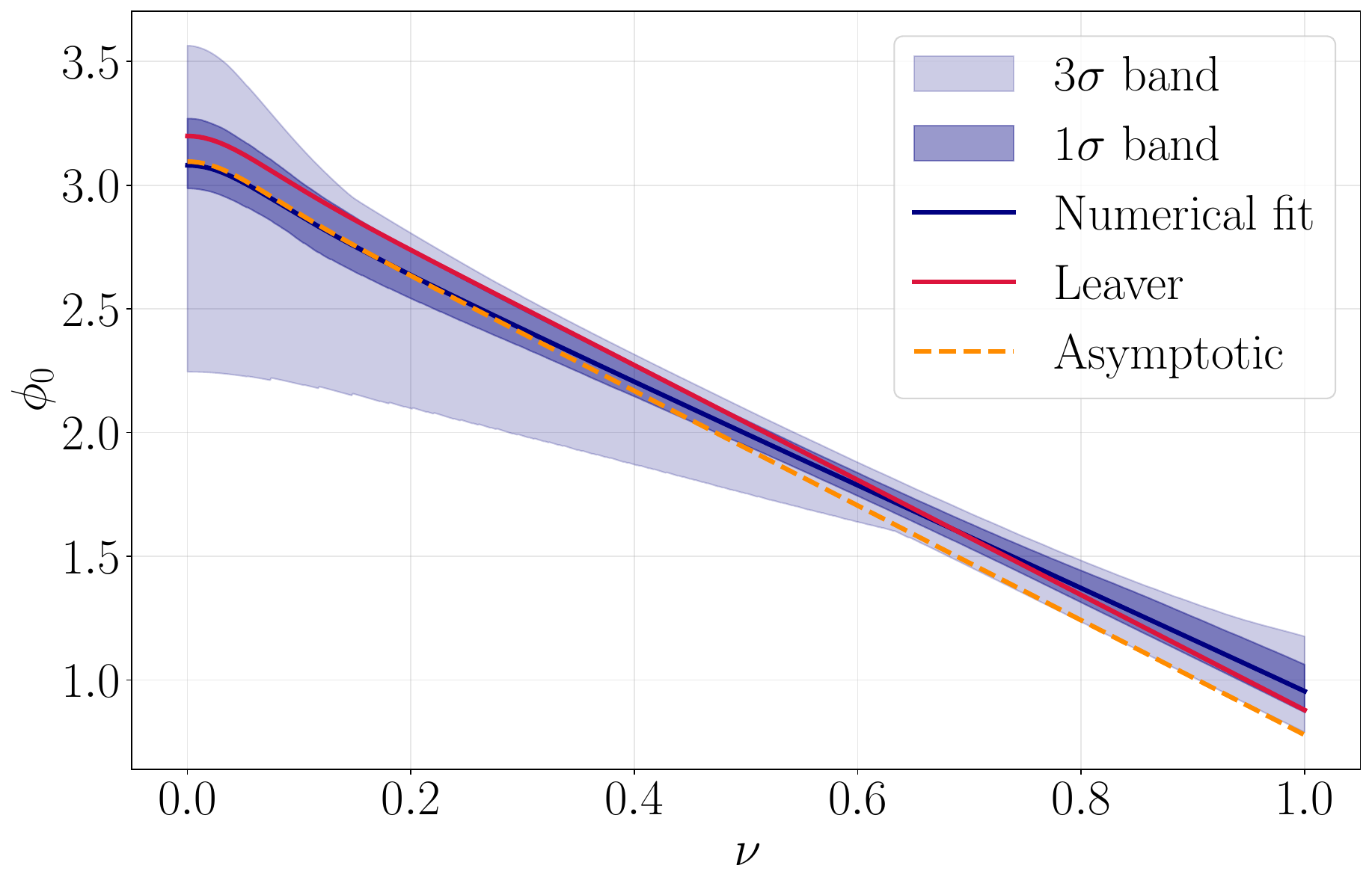}
}
\vspace{0.1cm}
\subfloat{%
\includegraphics[width=\linewidth]{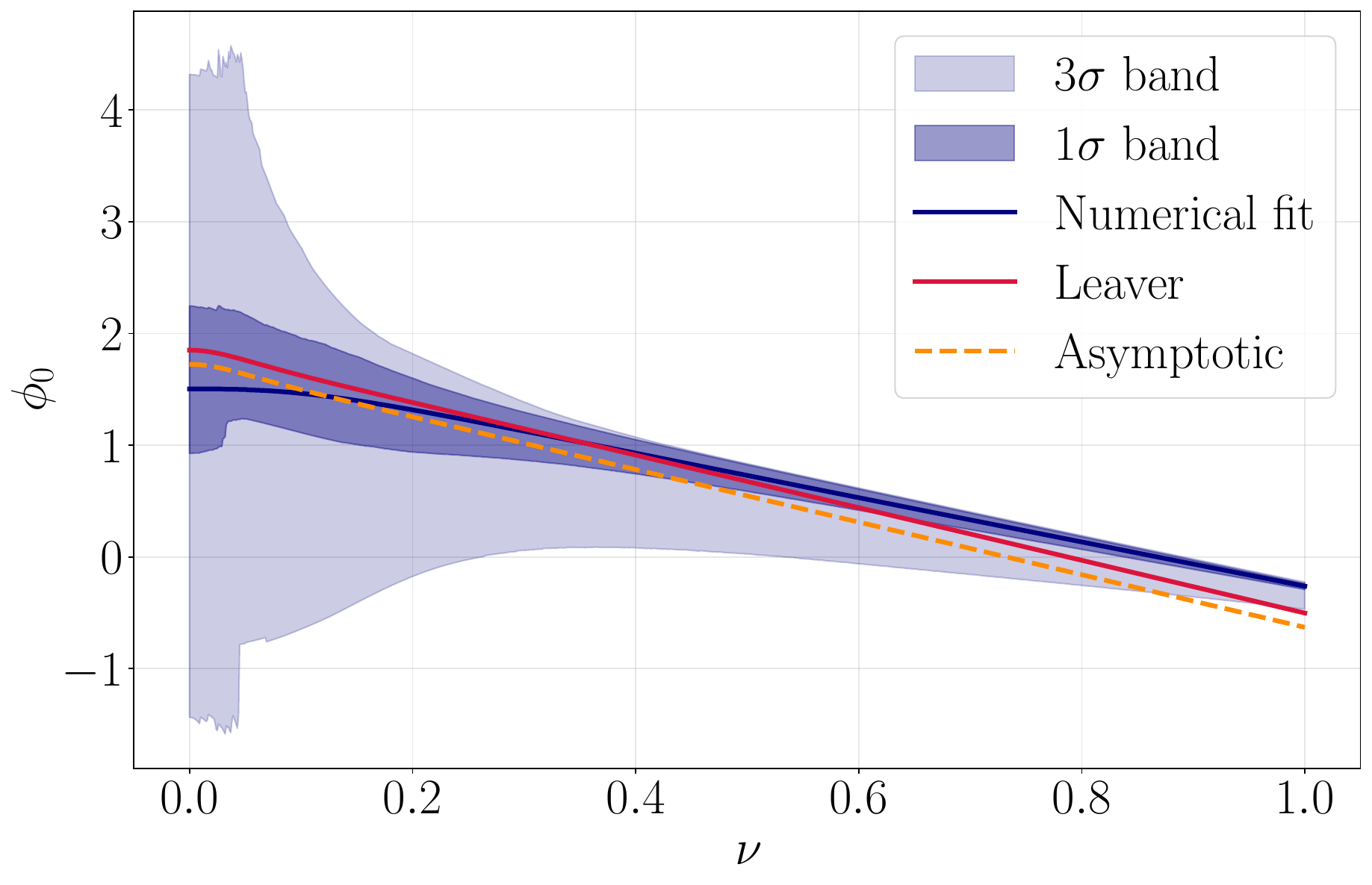}
}
\caption{\justifying 
Excitation phase of the fundamental mode $C_0$ as a function of the carrier frequency $\nu$ at fixed width $\sigma = 5$ for the higher multipoles \textbf{(top)} $\ell = 3$ and \textbf{(bottom)} $\ell = 4$. The values obtained from the Leaver and asymptotic methods are consistent with the $3\sigma$ confidence bands inferred from the fits. The free phase $\phi_0$ decreases monotonically with increasing $\nu$, while also exhibiting larger uncertainty bands for $\ell=3$ and $\ell=4$  with respect to $\ell=2$.}
\label{fig:C0_l3_l4_phase}
\end{figure}

Fig.~\ref{fig:C0_l3_l4_phase} shows the $\phi_0$ phase for the $\ell=3$ (top) and $\ell=4$ (bottom) modes in terms of the carrier frequency $\nu$. Notice that the trends are similar to these shown for $\ell=2$ (Fig.~\ref{fig:phases}). The Leaver, asymptotic, and numerical curves agree within the $3\sigma$ confidence bands. However, the uncertainty bands become broader, likely due to larger numerical errors affecting the higher modes.

\clearpage
\bibliography{main}

\end{document}